\documentclass[twocolumn,aps,prb,floatfix,10pt]{revtex4-2}
\usepackage{graphicx}
\usepackage{hyperref}
\usepackage{color}
\usepackage{amsmath}

\begin{document}
\title{
\texorpdfstring{Magnetic and electronic ordering phenomena in the [Ru$_2$O$_6$] honeycomb lattice compound AgRuO$_3$}
{Magnetic and electronic ordering phenomena in the Ru2O6 honeycomb lattice compound AgRuO3}}
\author{Walter Schnelle}
\email{walter.schnelle@cpfs.mpg.de}
\author{Beluvalli E.\ Prasad}
\altaffiliation[Now at ]{Department of Chemistry, RV Institute of Technology and Management, Bangalore, 560076, India}
\author{Claudia Felser}
\author{Martin Jansen}
\affiliation{Max Planck Institute for Chemical Physics of Solids, 01187 Dresden, Germany}
\author{Evgenia V.\ Komleva}
\author{Sergey V.\ Streltsov}
\affiliation{M.\ N.\ Miheev Institute of Metal Physics of Ural Branch of Russian Academy of Sciences, 620137 Ekaterinburg, Russia}
\affiliation{Ural Federal University, Mira St.\ 19, 620002 Ekaterinburg, Russia}
\author{Igor I.\ Mazin}
\affiliation{Department of Physics and Astronomy and Quantum Science and Engineering Center, 
George Mason University, 22030 Fairfax, Virginia, USA}
\author{Dmitry Khalyavin}
\author{Pascal Manuel}
\affiliation{ISIS Neutron and Muon Source, Rutherford Appleton Laboratory, Didcot OX11 0QX, U.K.}
\author{Sukanya Pal}
\author{D.\ V.\ S.\ Muthu}
\author{A. K.\ Sood}
\affiliation{Department of Physics, Indian Institute of Science, Bangalore 560012, India}
\author{Ekaterina S.\ Klyushina}
\author{Bella Lake}
\affiliation{Helmholtz Zentrum Berlin f\"ur Materialien und Energie, 14109 Berlin, Germany}
\affiliation{Institut f\"ur Festk\"orperphysik, Technische Universit\"at Berlin, 10623 Berlin, Germany}
\author{Jean-Christophe Orain}
\author{Hubertus Luetkens}
\affiliation{Laboratory for Muon-Spin Spectroscopy, Paul Scherrer Institute, 5232 Villigen PSI, Switzerland}
\date{\today}

\begin{abstract}
The silver ruthenium oxide AgRuO$_3$ consists of honeycomb
[Ru$_2^{5+}$O$_6^{2-}$] layers, and can be considered an analogue of SrRu$_2$O$_6$
with a different intercalation stage. We present measurements of magnetic
susceptibility and specific heat on AgRuO$_3$ single crystals which reveal a
sharp antiferromagnetic transition at 342(3)\,K. The electrical transport  
in single crystals of AgRuO$_3$ is determined by a combination of activated 
conduction over an intrinsic semiconducting gap of $\approx 100$\,meV and 
carriers trapped and thermally released from defects. From powder neutron
diffraction data a N\'eel-type antiferromagnetic structure with the Ru moments 
along the $c$ axis is derived. Raman and muon spin rotation spectroscopy
measurements on AgRuO$_3$ powder samples indicate a further weak phase 
transition or a crossover in the temperature range 125--200\,K. The transition 
does not show up in magnetic susceptibility and its origin is argued to be 
related to defects but cannot be fully clarified. The experimental findings are 
complemented by DFT-based electronic structure calculations. It is found that 
the magnetism in AgRuO$_3$ is similar to that of SrRu$_2$O$_6$, however with 
stronger intralayer and weaker interlayer magnetic exchange interactions.
\end{abstract}

\maketitle

\renewcommand{\textfraction}{0.0}
\renewcommand{\floatpagefraction}{1.0}
\setcounter{totalnumber}{8}
\setcounter{topnumber}{2}
\setcounter{bottomnumber}{2}


\section{Introduction}
\label{sec:intro}

Ruthenium and its compounds feature impressively diverse chemical and physical
phenomena. This is reflected, for instance, by the oxidation states accessible
within the wide span from $-2$ to $+8$ and by particular electronic and
magnetic ground states formed in molecular compounds as well as in extended
solids. This is especially relevant in condensed matter research, where
ruthenium oxides and chlorides continue to attract prominent attention.
Accessibility of different valence states of Ru results in a dramatic
variability of physical properties even within the small structural motif.
Indeed, honeycomb Ru$^{3+}L_3$ layers ($L$ = ligand atom) form one of the 
cleanest Kitaev system known so far, with strong bond-dependent anisotropic 
exchange \cite{BanerjeeA2017a,JohnsonRD2015a}, Ru$^{4+}L_3$ dimerizes at low
temperatures and forms a unique covalent bond liquid above 270\,$^{\circ}$C
\cite{KimberSAJ2014a}. In the compound SrRu$_2$O$_6$, Ru$^{5+}L_3$ was argued 
to feature unusual quasimolecular orbitals (QMO), which determine the 
suppression of the magnetic moment and robust antiferromagnetic (AFM) coupling
\cite{StreltsovS2018a,PchelkinaZV2016a}.

In different structural contexts Ru$^{4+}$ is noted for other unique physical
phenomenon. For example, the dioxide RuO$_2$ was recently shown to be an example
of a novel magnetic state of matter, an antiferromagnet without Kramers
degeneracy, with unique physical ramifications \cite{SmejkalL2020a}. It also
generates a series of perovskite compounds ranging from SrRuO$_3$, CaRuO$_3$
and BaRuO$_3$, all bad metallic, ferromagnetic or nearly-ferromagnetic, with a
strong effect of magnetic fluctuations on transport
\cite{KosterG2012a,CaoG1997a,MazinII1997a}, to Sr$_3$Ru$_2$O$_7$ featuring one
of the first experimentally observed magnetic quantum critical points, and
Sr$_2$RuO$_4,$ an enigmatic material for many years misidentified as
spin-triplet superconductor \cite{MackenzieAP2020a,PustogovA2019a}.

The recently synthesized AgRuO$_3$ \cite{PrasadBE2017a} (space group
$R\bar{3}c$, $a = 5.2261(6)$\,{\AA}, $c = 32.358(5)$\,{\AA}, $Z = 12$)
featuring Ru$^{5+}$ $4d^3$ species comes close to the conception of a 2D
material \cite{DuongDL2017a}, in as much it consists of stacked honeycomb
[Ru$_2$O$_6$] poly-oxoanions, where the empty octahedral sites are capped from
both sides with silver atoms. Thus, the resulting (Ag$_2$Ru$_2$O$_6$)$_n$
slabs are charge neutral and resemble giant molecules. Based on magnetic
susceptibility measurements on a powder sample and a preliminary evaluation of
time of flight (TOF) neutron diffraction data it was claimed that AgRuO$_3$
would show strong magnetic exchange coupling, however, no long range magnetic
order could be detected at that time \cite{PrasadBE2017a}, while in seeming 
contradiction with the experiment first-principle calculations were predicting
AgRuO$_3$ to be magnetically very similar to SrRu$_2$O$_6$ (supporting information 
for Ref.\ [12]). The latter has been investigated intensively -- experimentally 
and theoretically -- in the last years
\cite{HileyCI2014a,HileyCI2015a,SinghDJ2015a,StreltsovS2015a,HarikiA2017a,
OkamotoS2017a,PonosovYuS2019a,SuzukiH2019a,WangWangWang2015a,TianW2015a,StreltsovS2018a}
and was shown to have the transition to a N\'eel AFM groundstate at $\approx 560$\,K. 
Moreover, in 2019 also BaRu$_2$O$_6$ was synthesized which appears to have a 
related crystal structure \cite{MarchandierT2019a}.

Here, we present detailed measurements of the magnetic susceptibility, electrical 
resistivity, specific heat and Raman spectroscopy on AgRuO$_3$ single crystals and  
high-resolution TOF neutron data and muon spin rotation spectroscopy ($\mu$SR) data 
on polycrystalline material, all clearly demonstrating the presence of long-range
AFM order up to $\approx 342(3)$\,K. This relatively high N\'eel temperature
of AgRuO$_3$ is in agreement with our first principles analysis of the
electronic structure. In several respects AgRuO$_3$ parallels the properties
of SrRu$_2$O$_6$, however, there are conspicuous differences. Specifically,
there is a change of regime, possibly a second phase transition, around
125--200\,K within the AFM phase. It does not show up in either magnetic
susceptibility or specific heat, but there are clear changes in both $\mu$SR
and Raman spectra in this range, and, in roughly the same range, the resistance 
shows nontrivial and nonmonotonic temperature behavior which can be explained 
by defect states. 

In this paper we first present the extensive experimental findings, then
complement them with the electronic structure results, and finally sum up
relevant implications. While the N\'eel transition at 342\,K can be well
described in terms of quasi-2D magnetic interactions resulting from the
first principles calculation and QMOs, the nature of the low-temperature phase 
transition(s) or crossover is probably related to defect states, but remains 
not fully clarified.


\section{Experimental and calculation details}
\label{sec:exp}

For the measurements of various physical properties we used either portions
of, or selected sets of tiny single crystals from the large polycrystalline
sample of AgRuO$_3$ that had been synthesized as previously described
\cite{PrasadBE2017a}.

Magnetization was measured on a set of 10 selected single crystals (total mass
622\,$\mu$g) with a MPMS3 (SQUID-VSM, Quantum Design) magnetometer. The
crystals were glued with a minute amount of GE varnish to a quartz sample
holder with the $c$ axes oriented either parallel or perpendicular to the
applied field. Data were measured on warming after zero-field cooling (zfc)
and on cooling (fc) in different magnetic fields.

Electrical resistance was determined on several crystals with a four probe ac
method (RES option, PPMS9, Quantum Design). Contacts in the hexagonal $a$
plane of the crystals ($I\perp c$) or at the bases of short prismatic crystals
($I \parallel c$) were made with Pt wires (25\,$\mu$m) and silver paint. The
geometry factors could not be determined reliably (for resistivity data on a
polycrystalline sample see Fig.\ 2 of Ref.\ \cite{PrasadBE2017a}).

Heat capacity up to 300\,K was measured on a cold-pressed pellet with the HC
option of a PPMS9 (Quantum Design) in magnetic fields up to $\mu_0H = 9$\,T.
Above 300\,K the melting of the thermal contact agent (high vacuum grease
Apiezon N) lead to unreliable results. Therefore, additional heat capacity
data were obtained with a differential scanning calorimeter (DSC; PerkinElmer
DSC8500) at heating rates of 5 and 20\,K\,min$^{-1}$.

Neutron diffraction measurements were performed at the ISIS pulsed neutron and
muon facility of the Rutherford Appleton Laboratory (UK), on the
high-resolution cold-neutron diffractometer WISH located at the second target
station \cite{ChaponLC2011a}. Polycrystalline AgRuO$_3$ ($\approx 2$\,g) was
loaded into a cylindrical 6\,mm diameter vanadium can and measured on warming
using an Oxford Instruments cryostat and a CCR hot-stage, respectively.
Rietveld refinement of the crystal structure was performed using the FullProf
program \cite{Fullprof} against the data collected in detector banks at
average $2\theta$ values of 58$^{\circ}$, 90$^{\circ}$, 122$^{\circ}$, and
154$^{\circ}$, each covering 32$^{\circ}$ of the scattering plane.

Temperature dependent Raman spectroscopy experiments were carried out on
crystals using a Linkam THMS350V in the range of 77\,K to 385\,K. The sample
was cooled with a continuous flow of liquid nitrogen. Raman spectra were
recorded in a LabRam spectrometer (Horiba) in back-scattering geometry, using
a 50$\times$ objective and laser excitation of 532 and 660\,nm.

Zero-field (ZF) $\mu^+$SR ($\mu$SR) measurements of polycrystalline
AgRuO$_3$ were performed using Dolly and GPS muon spectrometers at the Swiss
Muon Source. For the measurements 1.8\,g of powder AgRuO$_3$ was pressed into
a pellet and mounted on a silver sample holder. The measurements were
performed over the temperature ranges 0.2\,K\,$< T <$\,65\,K using an Oxford
Heliox and 65\,K\,$< T <$\,295\,K in an Oxford Variox cryostat.

The crystal structure for the electronic structure calculations was taken from
Ref.\ \cite{PrasadBE2017a}. The band structure calculations were performed
using the Wien2k [nonmagnetic GGA (Generalized Gradient Approximation)],
including Wannier function projections \cite{WIEN2k} and Vienna Ab-initio
Simulation Package (VASP, all magnetic calculations) \cite{VASP}. We utilized
the projector augmented-wave (PAW) method \cite{Bloechl94} with the
Perdew-Burke-Ernzerhof (PBE) GGA functional \cite{PBE96}. The energy cutoff
chosen in VASP was $E_\mathrm{cutoff} \sim 600$\,eV and the $6\times6\times6$
Monkhorst-Pack grid of $k$ points was used in the calculations.


\section{Results}
\label{sec:results}

\subsection{Magnetism}
\label{sec:magnetism}

\begin{figure}[tbh]
\includegraphics[height=0.44\textwidth,angle=90]{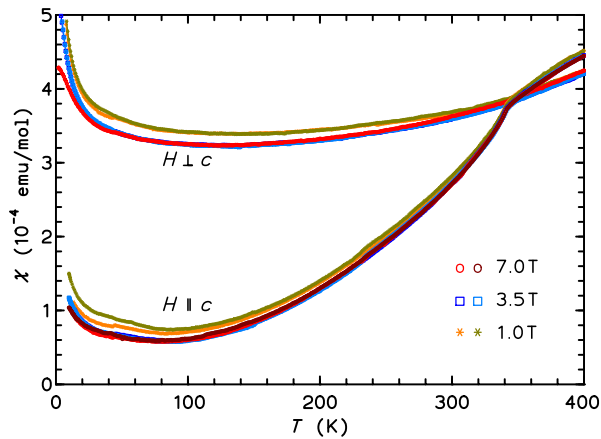}
\caption{Magnetic susceptibility $\chi(T)$ of AgRuO$_3$ crystals measured in
three different fields parallel and perpendicular to the crystallographic $c$
axes of the crystals. Light color symbols = zfc, dark color symbols = fc
protocol. Except for the lowest temperatures the zfc and fc curves coincide.
Also the data for 3.5\,T and 7.0\,T coincide over large temperature
intervals.}
\label{fig:susc}
\end{figure}

The magnetic susceptibility of AgRuO$_3$ up to 400\,K is shown in Fig.\
\ref{fig:susc}. Generally, the molar susceptibility is small. Above $T =
20$\,K it is only weakly dependent on the applied field and the zfc and fc
protocol data almost coincide. For $H \parallel c$ the susceptibility
$\chi_{\parallel}$ displays a sharp kink at $T_\mathrm{N} = 342$\,K, below
which $\chi_{\parallel}$ strongly decreases. $\chi_{\perp}(T)$ shows a weak
temperature dependence and a very weak kink at the same temperature (the slope
is larger above $T_\mathrm{N}$). Interestingly, the values of
$\chi_{\parallel}$ and $\chi_{\perp}$ coincide at $T_\mathrm{N}$ but at $T =
90$\,K the anisotropy $\chi_{\perp}/\chi_{\parallel} \approx 6$. These
findings suggest an AFM order with the Ru moments aligned along the $c$ axis.
Up to our maximum field of 7\,T the magnetization data show no indication for
a spin-flop or other metamagnetic transition.

Well below the ordering temperature, $\chi(T)$ for both directions show a weak
upturn, which we attribute to Curie-paramagnetic impurities or charged defects
in the sample. An estimate can be made by fitting a Curie law $C/T + \chi_0$
to the 1.0\,T data for $H\parallel c$ from 10--100\,K. $C$ corresponds to
0.10\,$\mu_\mathrm{B}$, equivalent to 0.3\,{\%} of spin 1/2 magnetic species.
It should be mentioned that in measurements on loose powder samples (cf.\
Fig.\ 3 in Ref.\ \cite{PrasadBE2017a}) preferential alignment of the
crystallites along the easy axis can occur at high applied fields (typically
$\mu_0H \geq 1$\,T). In measurements on powders in low applied fields no phase
transitions or indications for a weak ferromagnetic component of the AFM order
were detected.


\subsection{Electrical transport}
\label{sec:transport}

\begin{figure}[tbh]
\includegraphics[height=0.453\textwidth,angle=90]{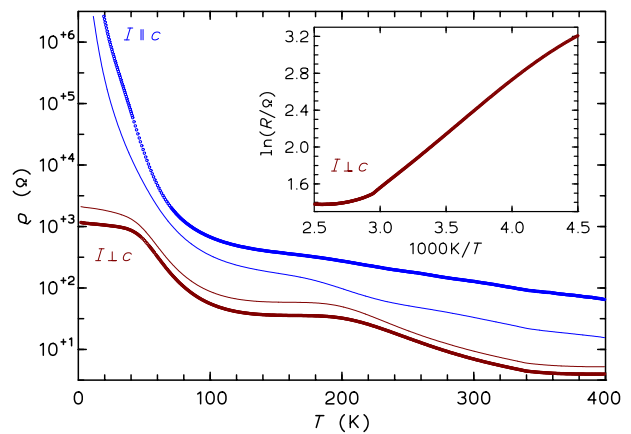}
\caption{Electrical resistance $R(T)$ of selected AgRuO$_3$ crystal for
current in the hexagonal $a$ plane ($I \perp c$) and current along the $c$
axis ($I \parallel c$). The inset shows an Arrhenius-type plot of the in-plane
resistance data. The thick and thin lines represent the data for two typical 
samples.}
\label{fig:rho}
\end{figure}

The temperature dependence of the electrical resistance was measured on
several single crystals. The characteristics of the resistance curves measured
with current in the $a$ plane ($I \perp c$) are well reproduced on several
crystals (Fig.\ \ref{fig:rho}, red curves). With decreasing temperature the
resistance generally increases and at $T_\mathrm{N} \approx 340$\,K a change
to a steeper slope (activated conduction, cf.\ inset of Fig.\ \ref{fig:rho})
can be observed. A plateau-like behavior is seen between $\approx 190$\,K and
$\approx 130$\,K, below which the resistance starts to rises again. Finally,
the in-plane resistance appears to reach another plateau below $T < 30$\,K.
This characteristic behavior is in stark contrast to the $c$ axis resistance
(Fig.\ \ref{fig:rho}, green curves), which is roughly one order of magnitude
higher and increases continuously with decreasing temperature, seemingly
indicating a different activated conduction behavior. At $T \approx 130$\,K a
changeover to a much stronger slope is observed. It has to be mentioned that
not all crystals showed such high resistance at low $T$ (compare the two green
curves in Fig.\ \ref{fig:rho}), which can be rationalized by the poor mechanical 
quality of the specimens. For all crystal a weak kink is visible in the $a$ 
plane resistance at the N\'eel temperature.

\begin{figure}[tbh]
\includegraphics[width = 0.453\textwidth]{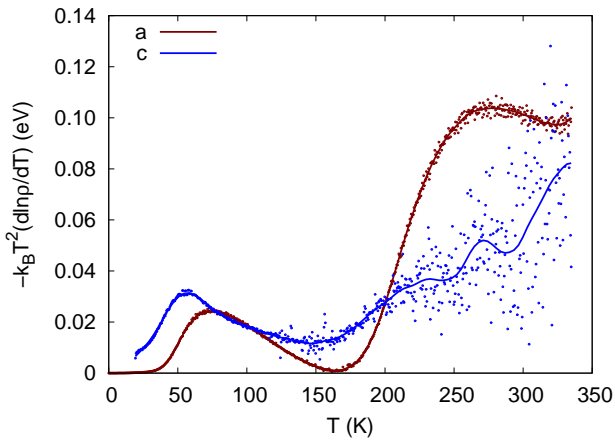}
\caption{Effective transport excitation gap defined as $E_\mathrm{g}(T) =
-k_\mathrm{B}T^2(d\ln\rho/dT)$ for current in the $a$ plane and along the $c$ 
axis. The continuous lines are spline-smoothed data.}
\label{fig:rhod}
\end{figure}

In order to get a better insight, we have calculated the function
$E_\mathrm{g}(T) = -k_\mathrm{B}T^2(d\ln\rho/dT)$ (Fig.\ \ref{fig:rhod}). For
both direction the activation barrier shows a profound temperature dependence,
by no means attributable to a single gap. The high-$T$ behavior is consistent 
with an activation gap of the order of 0.07--0.09\,eV for the both directions. 
The apparent gap drops nearly to zero around at $T \sim 150$\,K (170\,K), for 
the $c$ ($a$) directions, and raises again upon cooling to $\sim 55$\,K (75\,K) 
to the values of $\sim 0.035(0.025)$\,eV, respectively. Then it drops to zero 
again as $T \rightarrow 0$.

While the temperature dependence of the derivative-deduced activation energy
is qualitatively similar in both directions, the details are very different.
Since the actual gap is a scalar quantity, the only way to reconcile this
observation is to assume that there are several different reservoirs of the
current carriers, one of them corresponding to the actual excitation gap, with
should be larger that $\sim 0.1$\,eV, and the other(s) representing various
defect states inside the gap. A very small number of carriers are actually
metallic, and are responsible for the $T \rightarrow 0$ limit of the
activation energy. In addition, there are defect states that carry a small
number of carriers, and start to donate carriers at around 50\,K, but get
depleted at higher temperature. A closer inspection shows that one also needs
to introduce temperature dependent scattering, presumably related to the
same defect states and determined by a similar activation energy. Note that
the defect-derived carriers dominate the low-$T$ regime, while the high
temperature transport is effected by the carrier thermally excited across the
fundamental gap. With this in mind, we tried to fit the conductivity $\sigma 
= 1/\rho$ with:

\begin{align}
\sigma & = n\tau\\ n
       & = n_1 \exp(-E_{g}/T) + \frac{n_2}{(1+\exp(D/T)} \nonumber\\
1/\tau & = 1/\tau_0+\frac{1/\tau_1}{(1+\exp(D^{\prime}/T)} \nonumber
\label{eq:fit}
\end{align}

\begin{figure}[tbh]
\includegraphics[width = 0.405\textwidth]{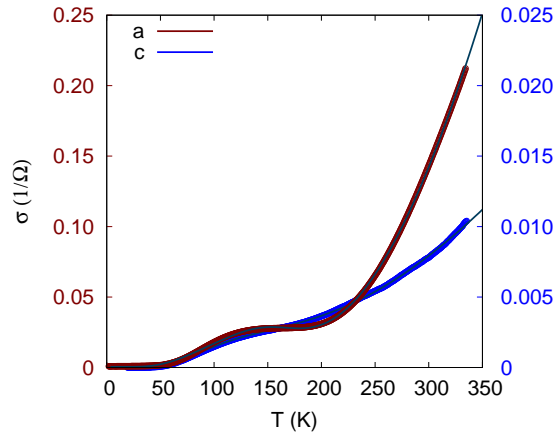}
\caption{Fitting to Eq.\ \ref{eq:fit} (lines) of the experimental conductivity
data (circles). The left (red) scale corresponds to the $a$-plane, the right 
(blue) scale to the $c$-axis conductivity.}
\label{fig:rhofit}
\end{figure}

As shown in Fig.\ \ref{fig:rhofit} this provides a perfect fit to the data,
although the parameters differ between the two directions. While the fit is not 
very sensitive to the parameters $D$ and $D^{\prime}$s and they can be forced 
to be the same, with only moderate loss of the fit quality, the fundamental gap
$E_\mathrm{g}$ is important for getting the high-temperature resistance
right. We can only conclude that the transport in the $c$ direction is
contaminated by grain boundaries and other defects, or is plainly not
Boltzmannian. The exceptionally large transport anisotropy is suggestive of
that. The formula (simplified in the sense that all defect states are lumped
together into a single level) is

\begin{align}
\sigma & = \frac{\sigma_0 + \sigma_1 \exp(-E_\mathrm{g}/T) +
\sigma_2/(1+\exp(D/T))} {1 + A/(1 + \exp(D^{\prime}/T))}
\end{align}.

The parameters derived from the $a$ plane transport are: $\sigma_0 = 5 \times
10^{-4}$, $\sigma_1 = 9 \times 10^3$, $\sigma_2 = 0.2$\,$\Omega^{-1}$,
$E_\mathrm{g} = 200$\,meV, $D = 20$\,meV, $D^{\prime} = 115$\,meV, $A = 1300$.
As mentioned, $D^{\prime}$ can actually be set to be the same as $D$ with only
moderate deterioration of the fit. For the $c$ axis these numbers are
$\sigma_0 = 9 \times 10^{-8}$, $\sigma_1 = 2.1$, $\sigma_2 = 0.08$
$\Omega^{-1}$, $E_\mathrm{g} = 110$\,meV, $D = 30$\,meV, $D^{\prime} =
30$\,meV, $A = 20$, but these numbers should be taken with a huge grain of salt.

One take-home message is that a defect level located at 20--30\,meV
(corresponding, roughly, to $T \sim 200$--300\,K) is capable of generating 
the anomaly in temperature dependence of transport, visually located at 150--170\,K.


\subsection{Specific heat}
\label{sec:cp}

\begin{figure}[tbh]
\includegraphics[width=0.48\textwidth,angle=0]{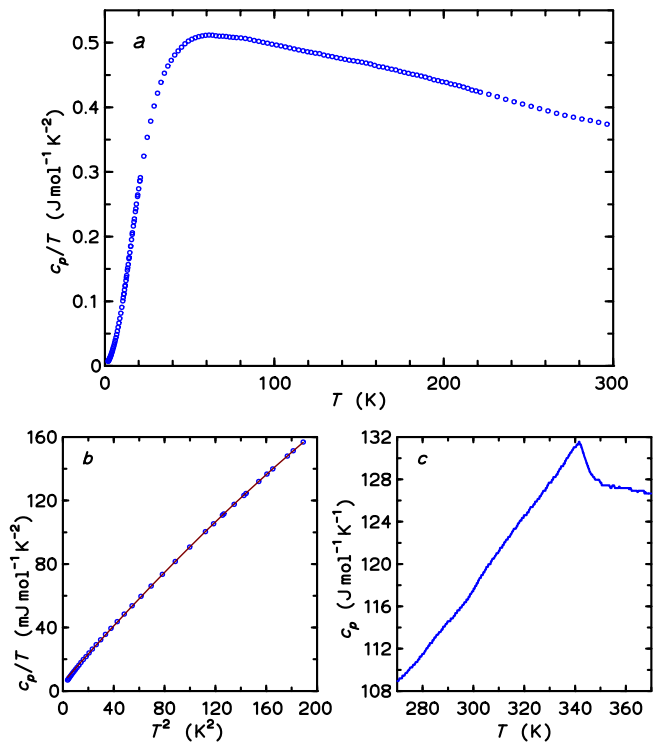}
\caption{(a) Specific heat of AgRuO$_3$. (b) low-temperature data in the
$c_p/T$ vs.\ $T^2$ representation. The line is the fit as detailed in the text
(c) high-temperature differential scanning calorimetry data (DSC; scan rate 
20\,K\,min$^{-1}$.}
\label{fig:cp}
\end{figure}

The specific heat is shown in Fig.\ \ref{fig:cp}a in a $c_p/T$ vs.\ $T$
representation. In the covered temperature range no peak or anomalous
broadened feature can be seen, suggesting the absence of any phase transition
in this temperature range. The data in the low temperature range (Fig.\
\ref{fig:cp}b) are well described by $c_p(T) = \gamma^{\prime}T + \beta T^3 +
\delta T^5$ with the latter two terms describing the phonon contribution. The
fit results in $\beta = 0.913(3)$ mJ mol$^{-1}$ K$^{-4}$, corresponding to the
initial Debye temperature of 220\,K, and $\delta = -0.00060(1)$ mJ mol$^{-1}$
K$^{-6}$ the next term in the harmonic lattice approximation constituting a
small correction. The linear coefficient $\gamma^{\prime} = 5.6(1)$ mJ
mol$^{-1}$ K$^{-2}$ is quite large (there should be no conduction
electron term). It is probably due to a rather high concentration of point
defects. It appears that the majority of these defects does not show up as an
upturn of the magnetic susceptibility towards low $T$ (cf.\ Fig.\
\ref{fig:susc}), thus they should be predominantly nonmagnetic, e.g.\ oxygen
or silver defects. The magnetic field dependence of $c_p(T)$ up to 9\,T is
very weak.

The DSC specific heat data (Fig.\ \ref{fig:cp}c) around room temperature are
about 5\,{\%} higher than the PPMS data and have realistic values close to the
Dulong-Petit limit ($c_p \approx 3nR$, where $n = 5$ is the number of atoms
and $R$ the gas constant). They display a clear second-order-type anomaly
with a midpoint at $T_\mathrm{N} \approx 344$\,K, corroborating the long-range
magnetic order at that temperature. No heating rate dependence of $c_p(T)$ or
latent heat is observed. We assign this transition to the AFM long-range 
ordering of the Ru moments.

\subsection{Neutron diffraction}
\label{sec:neutron}

\begin{figure}[tbh]
\includegraphics[width=0.48\textwidth]{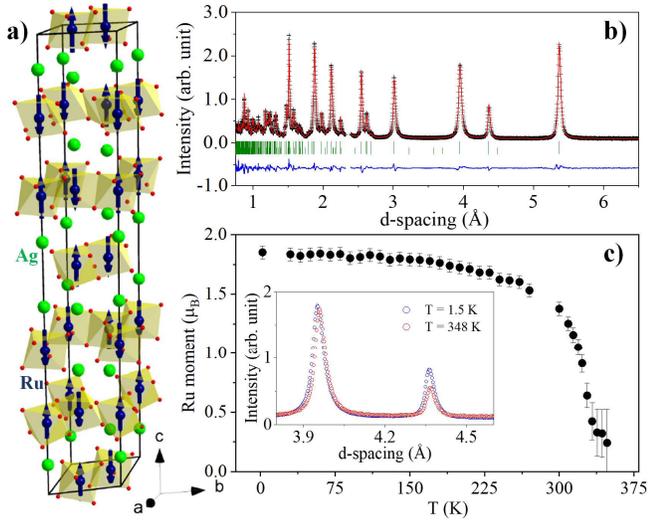}
\caption{(a) Schematic representation of the magnetic structure of AgRuO3 with
the $\bar{R}3^{\prime}c^{\prime}$ magnetic space group. (b) Rietveld
refinement of the neutron diffraction pattern collected at $T = 1.5$\,K
($R^\mathrm{Bragg}_\mathrm{nuclear} = 2.66$\,{\%} and
$R^\mathrm{Bragg}_\mathrm{magnetic} = 3.2$\,{\%}). (c) Ru ordered moment
versus temperature. The inset shows two peaks of the neutron diffraction
patterns collected at different temperatures.}
\label{fig:pnd}
\end{figure}

To further confirm the presence of long-range magnetic ordering, neutron
diffraction measurements (1.5\,K $ < T < $ 348\,K) were performed. The data
revealed that some of the reflections demonstrate a strong temperature
dependence below $T_\mathrm{N} \approx 335$\,K (Fig.\ \ref{fig:pnd}c, inset),
which is consistent with the fact that magnetic unit cell coincides with the
cell of the nuclear structure and the subsequent refinement of the diffraction
patterns (see, e.g., Fig.\ \ref{fig:pnd}b) was done assuming zero magnetic
propagation vector ($k = 0$).

The refinement was assisted by symmetry arguments based on the representation
theory \cite{ParkinsonNG2003a,Isotropy} and the parametrization for the
magnetic form factor of Ru$^{5+}$ has been taken from Ref.\
\cite{CampbellBJ2006a}. It has been found that a simple magnetic structure
with antiparallel alignment of spins on the nearest neighbor Ru sites (N\'eel
structure) provides a good refinement quality of the neutron diffraction data
(Fig.\ \ref{fig:pnd}b) in the whole temperature range below $T_\mathrm{N}$.
Surprisingly, the fitting quality was found to be almost insensitive to the
moment direction, complicating the choice between the models with in-plane and
out-of-plane spin polarization. However, our DFT calculations that take into
account spin-orbit coupling (see the Sec.\ \ref{sec:elstru} for details)
strongly support the model with out-of-plane spin orientation, In agreement
with the observed anisotropy of the magnetic susceptibility. Note that this
spin direction was also found experimentally in another trigonal ruthenate,
SrRu$_2$O$_6$ \cite{HileyCI2015a,TianW2015a}.

The magnetic structure of AgRuO$_3$ (Fig.\ \ref{fig:pnd}a) implies $\bar
{R}3^{\prime}c^{\prime}$ magnetic symmetry with the ordered moment size of
1.85(5)\,$\mu_\mathrm{B}$ at $T = 1.5$\,K. The moment varies smoothly with
temperature (Fig.\ \ref{fig:pnd}c) indicating that the anomalous behavior
found between 170 and 225\,K in the Raman and $\mu$SR measurements is likely
to be predominantly nonmagnetic in origin.


\subsection{Raman spectroscopy}
\label{sec:raman}

\begin{figure}[tbh]
\includegraphics[width=0.49\textwidth]{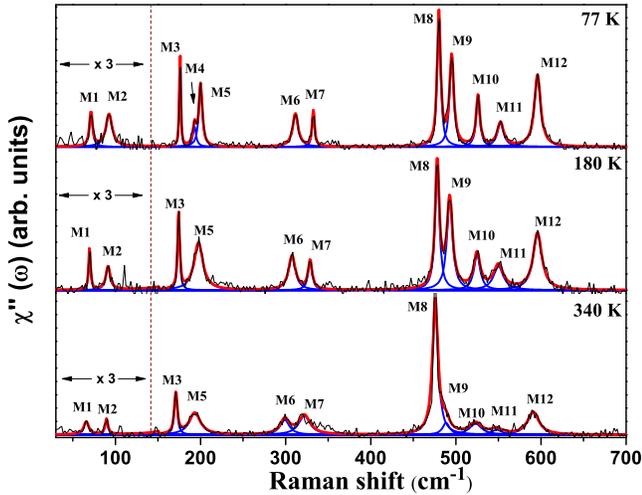}
\caption{Raman spectra of AgRuO$_3$ at typical temperatures (black lines). 
The solid red line is a Lorentzian fit to the data, blue solid lines are
individual fits of the Raman modes. The arrow indicates the additional 
mode M4.}
\label{fig:raman1}
\end{figure}

\begin{figure}[tbh]
\includegraphics[width=0.49\textwidth]{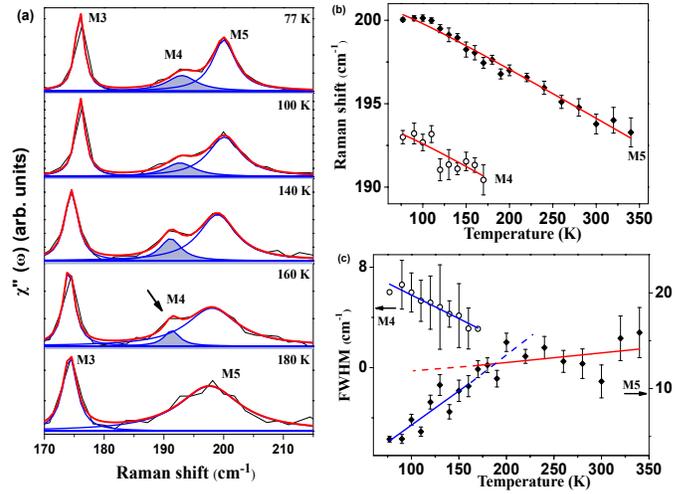}
\caption{(a) Temperature evolution of the Raman spectra in the range 
170--220\,cm$^{-1}$. The solid red line is a Lorentzian fit to the data, 
blue solid lines are individual fits of the Raman modes. The (b) Raman shift 
and (c) FWHM of the modes M4 and M5 plotted as a function of temperature. 
The red lines are fit to the anharmonic model and blue lines are linear fits 
to the data.}
\label{fig:raman2}
\end{figure}

Raman scattering has the ability to identify the role of phonons in structural
and magnetic transitions, the latter via spin-phonon coupling. At ambient
temperature AgRuO$_3$ has a trigonal $R\bar{3}c$ structure with 25 Raman
active modes of $A_{1g}$ and $E_g$ symmetries. Figure \ref{fig:raman1} shows
the reduced Raman susceptibility $\chi^{\prime\prime}(\omega) =
I(\omega)/(n(\omega)+1)$, where $I(\omega)$ is the observed intensity and
$[(n(\omega)+1)]$ is the Bose-Einstein factor, at a few typical temperatures.
Lorentzian line shapes were fitted to $\chi^{\prime\prime}(\omega)$ to extract
the phonon frequencies, full width at half maximum (FWHM), and integrated
intensities of the Raman modes. Table 1 in the Supplemental Material
\cite{AgRuO3-SuppMat} lists the experimentally (at 300\,K) observed and
calculated frequencies for the trigonal high-temperature (HT) AFM state, which
are shown with the corresponding error bars (for an explanation of how the
error bars were decided for the theoretical values, see Ref.\ 
\cite{AgRuO3-SuppMat}). The agreement is nearly perfect. At 77\,K, twelve
first-order Raman modes (M1 to M12) are observed in the spectral range 
30--650\,cm$^{-1}$. Most noticeable is the disappearance of mode M4 above 
170\,K, while all other modes are present in our working temperature range.

\begin{figure}[tbh]
\includegraphics[width=0.39\textwidth]{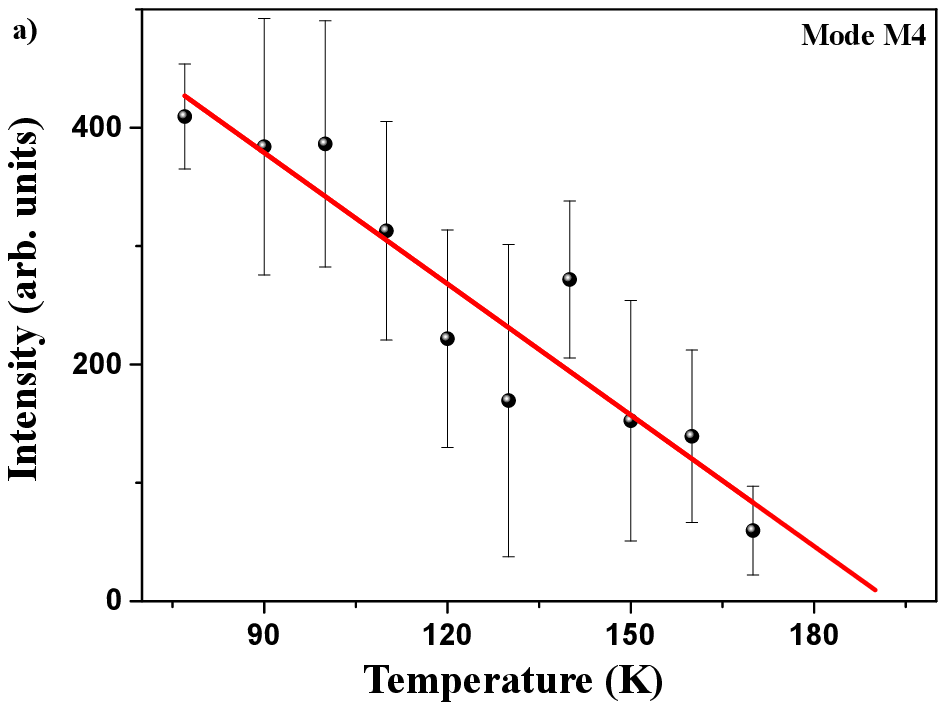}
\includegraphics[width=0.47\textwidth]{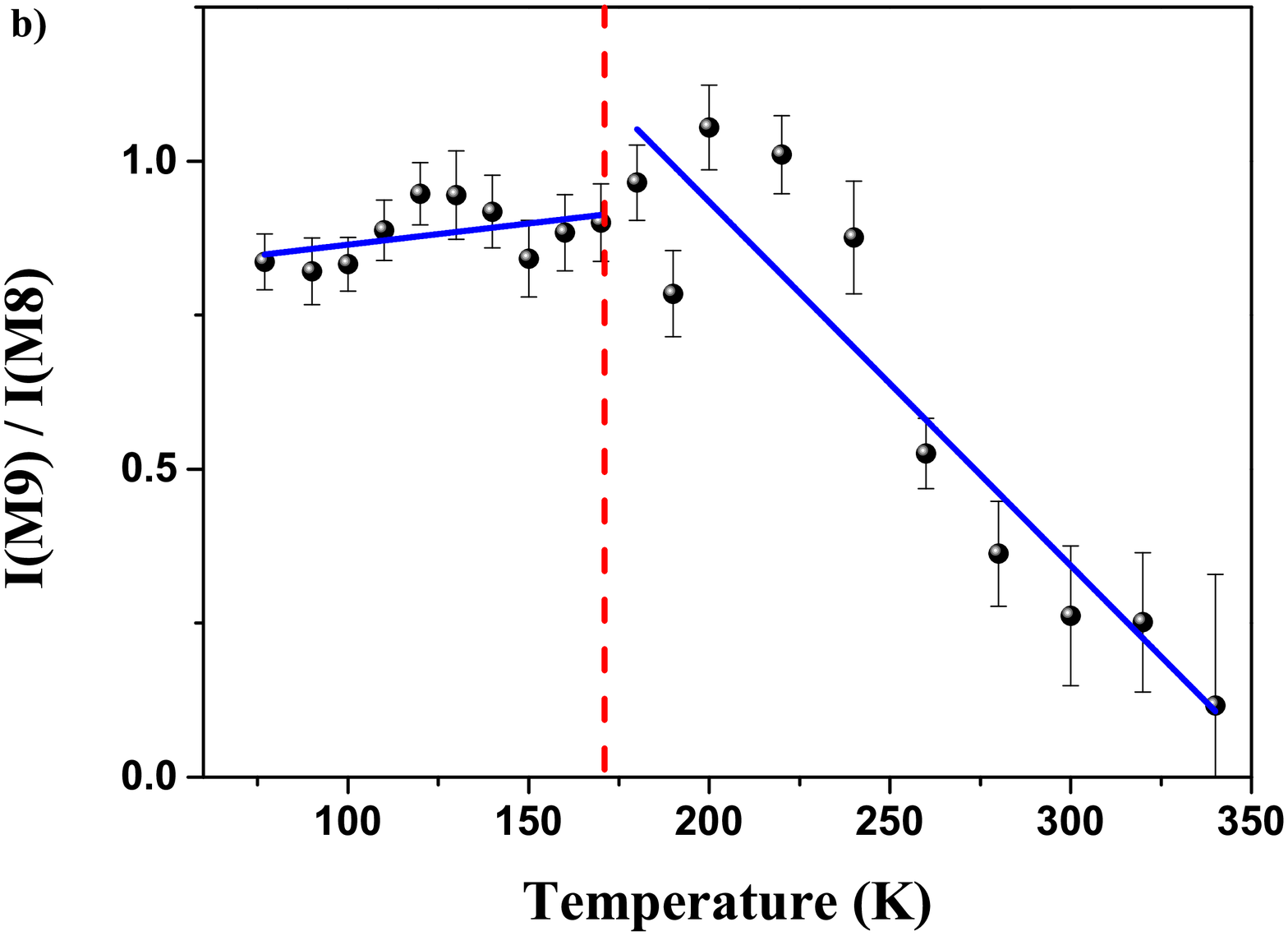}
\caption{(a) Temperature evolution of the integrated intensity of the M4 Raman
mode. The red line is a linear fit resulting in an intercept $T_\mathrm{c} =
193(5)$\,K. (b) The ratio of integrated intensities of mode M9 to M8 shows a
change in slope around 170\,K. The blue solid lines are linear fits to the
data below and above 170\,K.}
\label{fig:raman5}
\end{figure}

To bring home this point clearly, Fig.\ \ref{fig:raman2}a presents the fitted
Raman spectra for selected temperatures in the spectral range 170--225
cm$^{-1}$. The temperature evolution of the mode frequencies and FWHM for
modes M4 and M5 are shown in Fig.\ \ref{fig:raman2}b and \ref{fig:raman2}c,
respectively. The FWHM of the mode M4 is anomalous, i.e.\ it increases as
temperature is lowered. The FWHM of the mode M5 also shows a significant
change at 170\,K. The temperature dependence of the frequencies and linewidths
of the remaining phonons is given in the Supplemental Material, Fig.\ 1 and
Fig.\ 2 \cite{AgRuO3-SuppMat}, respectively. Phonon frequencies show the
expected behavior of decreasing frequency as $T$ is increased due to
quasi-harmonic (i.e.\ thermal expansion effect) and anharmonic effects
\cite{KlemensPG1966a}. The solid red lines are fits to the simple cubic
anharmonic model where a phonon decays into two phonons of equal frequencies
\cite{KlemensPG1966a}, given by $\omega^\mathrm{cubic}(T) = \omega
(0)+C[1+2n(\omega(0)/2)]$, $C$ being the self-energy parameter. Coming back to
mode M4, it is clearly visible for both polarizations at $T = 77$\,K (see Fig.\ 
3 of the Supplemental Material \cite{AgRuO3-SuppMat}) but disappears at around 
170\,K. Fig.\ \ref{fig:raman5}a shows the integrated intensity of this mode as 
a function of $T$. Given the error bars, this dependence is equally well 
consistent with a linear behavior (shown), with an exponential decay in the 
entire temperature range, or a square-root dependence corresponding to a second 
order phase transition at $T_\mathrm{c} = 170$\,K (both not shown).

Further, as shown in Fig.\ \ref{fig:raman5}b, the intensity of mode M9
normalized with respect to mode M8 shows a distinct change at 170\,K. The
intensity of M9 gradually decreases with $T$ and goes to zero near
the AFM ordering temperature $T_\mathrm{N}$. All these observations of
distinct changes in temperature dependence of the phonon modes M4, M5 and M9
indicate subtle structural changes around $T\approx 170$\,K. However, absence
of any distinct feature at this temperature in other probes suggest that these
changes do not constitute a true phase transition, but indicate a crossover
region probably stretching between 150 and 200\,K. Note that this is the same
range where the in-gap defect states strongly affect electrical transport,
albeit a mechanism by which free carriers donated by these states can affect
the Raman scattering is not clear.

\begin{figure}[tbh]
\includegraphics[width=0.41\textwidth]{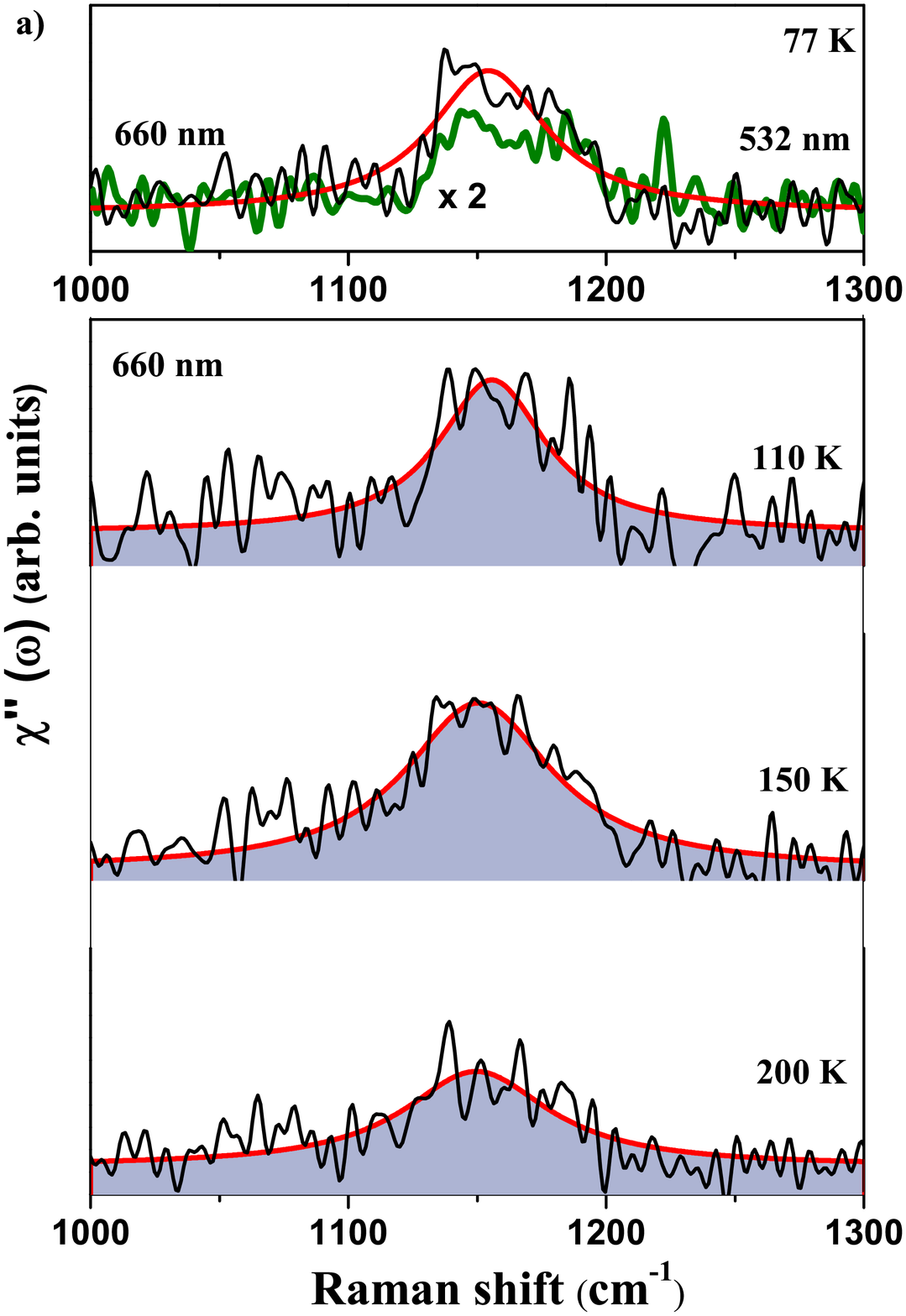}
\includegraphics[width=0.45\textwidth]{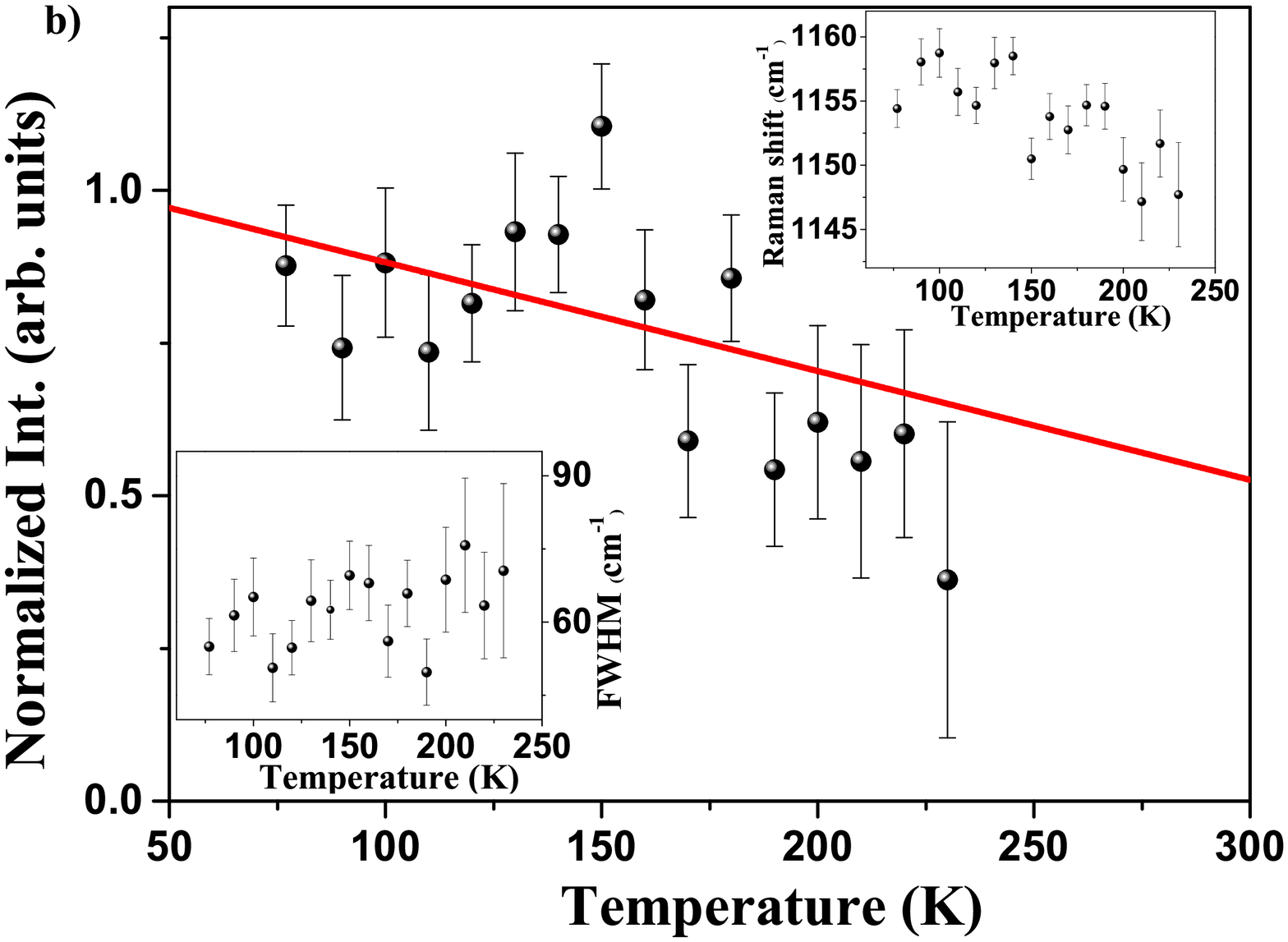}
\caption{(a) The broad mode is observed with both 532\,nm (shown by solid thick
green line) and 600\,nm lasers at 77\,K. Temperature evolution of the broad mode
of with 660\,nm red laser is shown subsequently. Raw data (high frequency noise
has been filtered out). The red lines are Lorentzian lineshape fits. 
(b) Temperature dependence of the intensity of the broad mode. The red solid 
line is a linear fit to the data. The frequency and the linewidth as a function 
of temperature are shown in the insets.}
\label{fig:raman6}
\end{figure}

We now come to the higher spectral range shown in Fig.\ \ref{fig:raman6}a
where we observe a weak broad band at 1155 cm$^{-1}$. To establish that this
is a Raman mode and not photoluminescence, we did experiments at 77\,K with
two different incident photon excitations, 660\,nm and 532\,nm (shown with the
thick green line in Fig.\ \ref{fig:raman6}a). This confirms that the mode at
$\approx 1155$ cm$^{-1}$ is a Raman mode. Fig.\ \ref{fig:raman6}b shows the
temperature dependence of the integrated intensity of this broad band. The red 
line (a linear fit) is intended as a guide to the eye.

The $T$-dependence of the frequency and the linewidth are shown in the insets 
of Fig.\ \ref{fig:raman6}b. In several Heisenberg antiferromagnets, like in 
YBa$_2$Cu$_3$O$_6$ \cite{KnollP1990a}, Sr$_2$IrO$_4$, Sr$_3$Ir$_2$O$_7$ 
\cite{GretarsonH2016a}, and SrRu$_2$O$_6$ \cite{PonosovYuS2019a} two-magnon 
Raman modes have been observed in the magnetically ordered state. If the 
Heisenberg model is defined as

\begin{equation}
H = \sum_{i \neq j} J_{ij} \mathbf{S}_i \mathbf{S}_j,
\label{eq:Ham}
\end{equation}

then the frequency of two-magnon Raman scattering is $\sim
17J$.\cite{PonosovYuS2019a}. Taking $s = 1$ we get an estimate of $J \approx
97$\,K, close to our calculated value (see Sec.\ \ref{sec:elstru} below).

However, we cannot completely rule out the possibility of electronic Raman
scattering associated with the electronic transition between the
QMOs of Ru$^{5+}$, as in the case of SrRu$_2$O$_6$ \cite{PchelkinaZV2016a}, 
although, in principle, one would expect such transitions to manifest at 
higher energies.


\subsection{Muon spin rotation spectroscopy}
\label{sec:muSR}

\begin{figure}[ptb]
\includegraphics[width=0.48\textwidth]{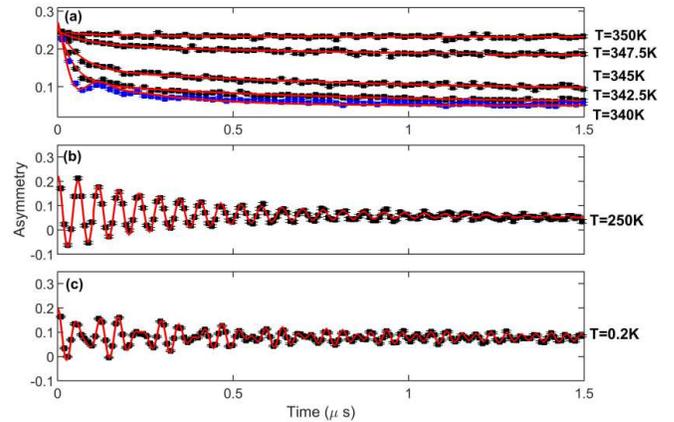}
\caption{Zero-field $\mu$SR spectra measured on powder AgRuO$_3$ at (a) $T =
342.5$, 345, 347.5 and 350\,K using the GPS muon spectrometer; (b) $T = 250$
and (c) $T = 0.2$\,K using the Dolly spectrometer. The red lines represent the
best results of the fit analysis described in the text.}
\label{fig:musr1}
\end{figure}

\begin{figure}[ptb]
\includegraphics[width=0.48\textwidth]{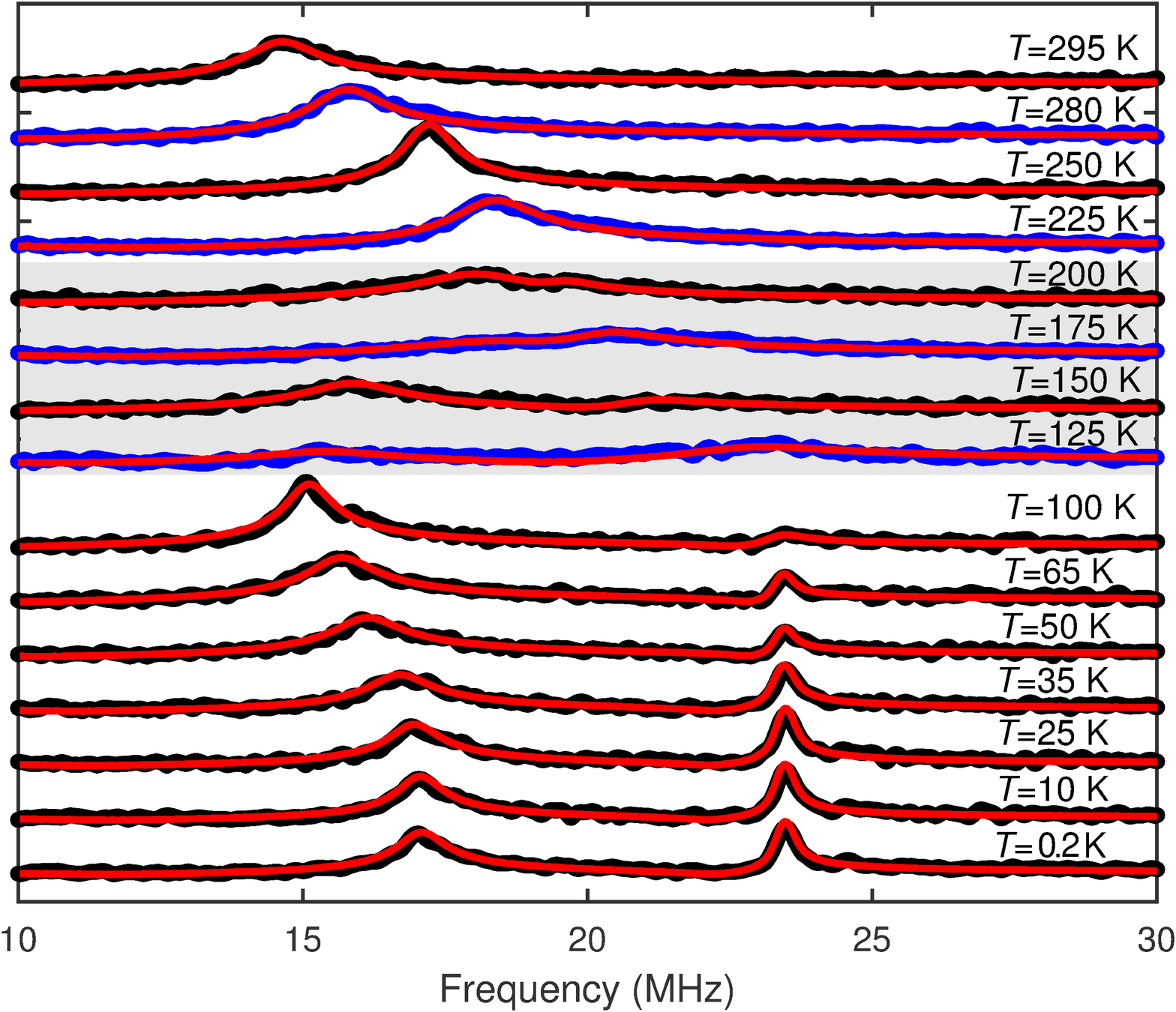}
\caption{Fast Fourier transformation of the muon spin-precession signal measured 
in zero-field at various temperatures using the Dolly (black) and GPS (blue) 
spectrometers. The FFt of the best fit to Eq.\ \ref{eq:fitfun1} is given
by the red lines through the FFt data. The gray area indicate the temperature 
regime 125\,K $\leq T \leq$ 200\,K.}
\label{fig:musr2}
\end{figure}

Figure \ref{fig:musr1}c shows the time-dependence of the muon spin
polarization in AgRuO$_3$ measured on the DOLLY spectrometer at the 
base temperature of $T = 0.2$\,K in zero field (ZF). The modulated 
oscillations indicate the presence of an internal magnetic field in the sample 
due to long-range magnetic order. The Fast Fourier transformation (FFt) of the 
data (Fig.\ \ref{fig:musr2}) shows that, in general, there are two fractions of 
the muon spin ensemble which oscillate with different frequencies giving rise to two
peaks $\approx 17$ and $\approx 23$\,MHz. The presence of two frequencies can
be attributed to two magnetically and/or crystallographically inequivalent
muon stopping sites in the crystal structure. To extract the frequencies of
the observed oscillations, the ZF-$\mu$SR spectrum at $T = 0.2$\,K was
analyzed with the software package MUSRFIT \cite{MUSRFIT} using the
fitting function:
\begin{equation}
A(t) = \sum_{i = 1}^2A_i\cos(2\pi\nu_it)\exp(-\lambda_it)+A_3\exp(-\lambda_3t) \label{eq:fitfun1}
\end{equation}
Here, $A_1$ and $A_2$ are the amplitudes of the muon spin oscillations due to
the component of the internal field which is perpendicular to the muon spin
polarization; $\nu_1$ and $\nu_2$ are the frequencies of these oscillations
which are connected to the internal field via the relation $\nu_i =
\gamma_{\mu}|B_i|/2\pi$ where $\gamma_{\mu}$ is the muon gyromagnetic ratio
and $|B_i|$ are the averaged internal fields on the two stopping sites. The
exponential terms describe the damping of the oscillation with relaxation
rates $\lambda_1$ and $\lambda_2$ and assume a Lorentzian field distribution.
The amplitude $A_3$ and the corresponding relaxation rate $\lambda_3$ take
into account a contribution of nonoscillatory signal due to the interaction of
the muons with the internal fields which are parallel to the initial muon spin
polarization. The best fit was achieved for the frequencies of $\nu_1 =
17.074 \pm 0.006$\,MHz and $\nu_2 = 23.456\pm0.003$\,MHz at $T = 0.2$\,K which
correspond to local fields of 1.26\,T and 1.3\,T respectively. Furthermore, the
harmonic form of the oscillations indicates that the long-range magnetic order
in the ground state of AgRuO$_3$ is commensurate.

\begin{figure}[ptb]
\includegraphics[width=0.44\textwidth]{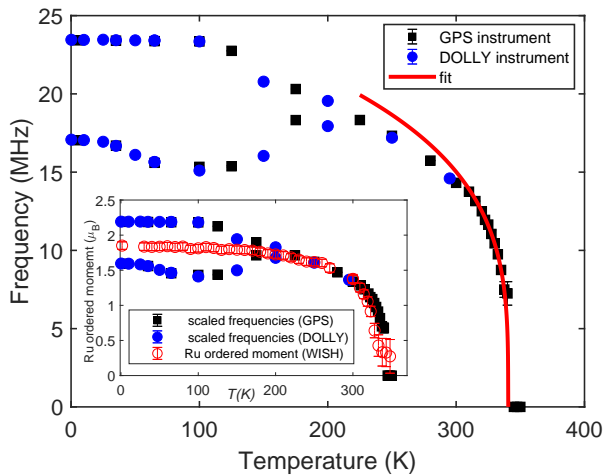}
\caption{Temperature dependence of the muon-spin precession frequencies
extracted from the analysis of ZF-$\mu$SR spectra of AgRuO$_3$. Blue 
circles and black squares correspond to the frequencies extracted from
the muon data collected at the Dolly and GPS spectrometers, respectively.
The red line gives the best fit achieved using Eq.\ \ref{eq:fitcrit}.
The inset shows a comparison of the scaled ZF-$\mu$SR frequencies (blue 
circles and black squares) with the Ru ordered moment (red circles) from 
neutron diffraction data.}
\label{fig:musr3}
\end{figure}

Further ZF-$\mu$SR measurements were performed  using DOLLY and GPS spectrometers 
to explore the temperature
dependence of the harmonic modes observed at base temperature. Figure
\ref{fig:musr2} shows the evolution of the FFt of ZF-$\mu^+$SR spectra
collected up to 295\,K. The frequencies of the oscillations in the
ZF-$\mu^+$SR spectra correspond to the peaks in the FFt. The solid red lines
represent the FFt of the best fit of the data achieved using Eq.\ \ref{eq:fitfun1}. 
As Fig.\ \ref{fig:musr2} shows, there are three different temperature regimes:
(i) $T < 125$\,K, (ii) 125\,K $\leq T \leq$ 200\,K and (iii) $T > 200$\,K. 
Below $\sim 125$\,K, there are two distinct and well-defined frequencies, the lower 
one noticeably softening with temperature, from $\approx 17$\,MHz at $T = 0.2$\,K 
to $\approx 15$\,MHz at $T = 100$\,K. The other one is basically $T$-independent, 
at $\nu\approx 23.5$\,MHz. Between $T = 125$ and 200\,K the spectra are extremely
broad with some traces of one or possibly two frequencies within the range
$\nu\approx 16$\,MHz and $\nu\approx 25$\,MHz. Furthermore, the relative
intensity and widths (but not frequencies!) of the spectral features in this
range is very sample-dependent (note that the sample used on the two different
instruments were from different batches), indicating a possible role of
impurities. A single sharp frequency which is sample independent reappears at
$\approx 17$\,MHz for $T = 225$\,K, and is clearly visible at $T = 295$\,K, at
$\approx 15$\,MHz. The extracted frequencies are plotted as a function of
temperature in Fig.\ \ref{fig:musr3}. The points between 125\,K and 200\,K
where the FFt of the signal is unclear give our the best guesses of the possible
position of two peaks in FFt while below 125\,K two clear frequencies are
observed.

Intriguing, the enigmatic intermediate temperature regime spans the same
region where the Raman spectra undergo qualitative changes and the
differential resistance is positive. It is possible that all three phenomena
have the same origin. The resistance analysis and the temperature-selective
$\mu$SR spectra suggest that charge dynamics related to defect traps may be
relevant.

To explore the higher temperature regime, where a single frequency was
observed, the ZF-$\mu$SR spectra of AgRuO$_3$ were collected on the GPS
spectrometer over the temperature range 300\,K $< T <$ 350\,K with step in
temperature of 5\,K and 2.5\,K. The data were analyzed using the function:
\begin{align}
A(t)  &   = F_\mathrm{m}\left[  A_0\cos(2\pi\nu_0t)\exp(-\lambda
_0t)+A_3\exp(-\lambda_3t)\right] \nonumber\\
&  +(1-F_\mathrm{m})G_\mathrm{KT} \label{eq:fitfun2}
\end{align}
where $F_\mathrm{m}$ is the fraction of the magnetically ordered state. This
function includes a Kubo-Toyabe term $G_\mathrm{KT}$ which takes into account the
contribution of the nuclear spins in the paramagnetic state of the electronic
spin system which was neglected at lower $T$. The results reveal that $T = 
340$\,K is the highest temperature where the oscillations exist in 
ZF-$\mu^+$SR data of AgRuO$_3$. Indeed, the $\mu^+$SR spectrum
at $T = 340$\,K (blue squares in Fig.\ \ref{fig:musr1}a) displays a weak
oscillation with frequency $\nu_0 = 7.25 \pm 0.75$\,MHz extracted from the fit
(red line). The FFt and fit analysis of the data at $T = 342.5$, 345, 347.5
and 350\,K reveal no oscillations and are well reproduced by using only
Kubo-Toyabe and nonoscillating terms. The depolarization of the asymmetry
decreases progressively in the ZF-$\mu^+$SR data at $T = 342.5$, 345 and
347.5\,K, suggesting a volume-wise destruction of the magnetic state with
increasing magnetic disorder. The muon polarization is fully recovered at $T =
350$\,K.

\begin{figure}[ptb]
\includegraphics[width=0.44\textwidth]{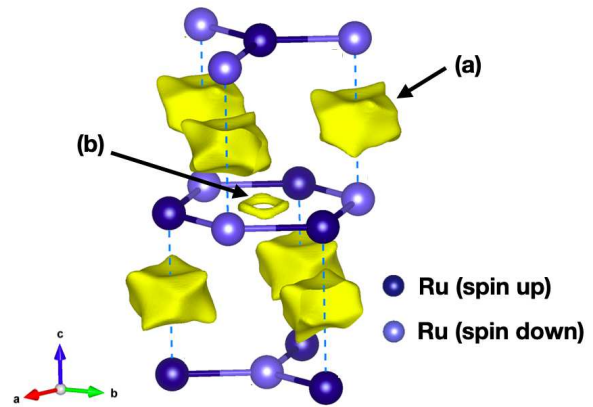}
\caption{Electrostatic potential in the antiferromagnetically ordered AgRuO$_3$, 
showing the potential muon trapping sites (yellow). The Ru atoms (purple) and 
some outlines of the $R\bar{3}c$ crystallographic unit cell are shown.}
\label{fig:musrpot}
\end{figure}

The extracted frequencies are plotted as a function of temperature in Fig.\ 
\ref{fig:musr3}. The observed temperature dependence can be compared with that 
of the Ru ordered moment extracted from neutron diffraction measurements because the 
frequencies are proportional to the local fields. The inset of Fig.\ \ref{fig:musr3} 
shows the scaled frequencies plotted over the Ru ordered moment. The curves were 
matched at $T = 250$\,K since this temperature is far from the critical region 
(vicinity of $T_\mathrm{N}$) where neutrons and muons have different sensitivity due 
to the different time scales of the techniques. The comparison reveals a general good 
agreement of the temperature dependencies. The deviation above $\approx 310$\,K can be 
attributed to the fact that the ordered moment in a $k = 0$ magnetic structure may be 
underestimated in a fit analysis in the vicinity of $T_\mathrm{N}$ where the signal is weak. 

To determine the N\'eel temperature $T_\mathrm{N}$ more accurately, and to
address the issue of the critical dynamics, the temperature dependence of
$\nu_0$ was fitted by:
\begin{equation}
{\nu(t)\propto(1-(T/T_\mathrm{N}))^\beta} \label{eq:fitcrit}
\end{equation}
where $\beta$ is the critical exponent and the range of fit was
325\,K $< T <$ 337.5\,K. The resulting value $T_\mathrm{N} =
340.5\pm0.5$\,K is in excellent agreement with the N\'eel temperature
revealed by magnetic susceptibility and other methods. However, the critical
exponent $\beta = 0.27\pm0.02$ is considerably reduced from the conventional
powers $\beta$ = 0.312 \cite{BramwellST1986a}, $\beta$ = 0.3485
\cite{CampostriniM2001a} and $\beta = 0.36$ \cite{BramwellST1986a} derived for
the 3D Ising, 3D XY and 3D Heisenberg magnetic systems, respectively. Such a
reduction, of various strength, can be attributed to the dominance of the
two-dimensional rather than three-dimensional correlations and is often
encountered in quasi-2D layered magnets (e.g.\ see Table A.1 in Ref.\
\cite{TaroniA2008a}). A possible explanation is the presence of long-range
interactions which can affect and reduce the observed critical exponent from
the theoretical predictions \cite{FisherME1972a}. In particular, according to
Ref.\ \cite{FisherME1972a}, $\beta = 0.27$ would correspond to a 2D magnet
with a long-range interaction decaying as $\sim1/r^{3.3}$. The presence of
long-range interactions is consistent with the quasimolecular picture
suggested by theory (see Sec.\ \ref{sec:elstru}), since such orbitals
typically generate long-range magnetic interactions \cite{WinterSM2017a}.

In order to elucidate the possible nature of the observed features we have
calculated the electrostatic potential in the $R\bar{3}c$ structure 
(Fig.\ \ref{fig:musrpot}). The energy landscape is 
quite uniform, with only two minima sufficiently deep to trap muons. They are 
located at the $6b$ and $6a$ positions, with rhombohedral coordinates $(0,0,0)$ 
and $(1/4,1/4,1/4)$ at the centers of the Ru$_6$ hexagons and in the middle between 
the vertical Ru-Ru bonds, respectively. In the neutron-determined magnetic 
$\bar{R}3^{\prime}c^{\prime}$ structure both sites have zero magnetic dipole field 
by symmetry. In reality, of course, the muons are shifted toward oxygens, and the 
calculation shows that stopping sites are $\approx 1.2$\,{\AA} removed from the 
oxygens. For each of the two positions there are 6 equivalent stopping sites, 
corresponding to the number of surrounding oxygens.

At high temperature the muons might be able to hop among these six position;
if this hopping rate is of the same order as the precession frequency, $\mu$SR
signal will be greatly broadened and when the hopping rate becomes larger, the
observed frequency will be reduce to zero, since the local fields averaged
over six stopping sites cancels by symmetry. Possibly, this explains our
observations at 125--200\,K. However, the origin of the single mode observed at
higher temperature is unclear at the moment. Indeed, neutron scattering does
not reveal anything unusual in terms of the long-range magnetic moment at
these temperatures. Moreover, our calculations suggest that the potential
barrier between (1/4,1/4,1/4) and (0,0,0) stopping sites is large enough to
prevent muon hopping between them at this temperatures. One possibility,
suggested by the transport measurements discussed above, is that defect states
may play a role. If at some characteristic temperature a particular kind of
defects (say, Ru vacancies) change they charge state, they may provide new
traps for muons, thus generating a new mode. However we do not have any direct
experimental or theoretical evidence for this scenario. It worth mentioning
that similar frequency splitting was previously observed in other quasi-2D
magnets Sr$_2$CuO$_2$Cl$_2$ \cite{LeLP1990a} and Ca$_{0.86}$Sr$_{0.14}$CuO$_2$
\cite{KerenA1993a} where no related phase transitions were detected using
neutron diffraction.


\subsection{Electronic structure}
\label{sec:elstru}

We start with simple GGA calculations. Similar to SrRu$_2$O$_6$, it is only
possible to achieve converge for a N\'eel AFM configuration (all nearest
neighbors must be AFM ordered) \cite{SinghDJ2015a,StreltsovS2015a}. Other
solutions such as ferromagnetic (FM), zigzag-AFM, or stripy-AFM do not survive
and collapse to the nonmagnetic configuration. This demonstrates that
AgRuO$_3$ cannot be described by a conventional Heisenberg Hamiltonian, due to
the itinerant nature of its electronic structure. The magnetic moment on the
Ru ion is 1.24\,$\mu_\mathrm{B}$, roughly consistent with experiment. Similar 
to SrRu$_2$O$_6$ this reduction is not related to any covalency effects, but can 
be traced down to the formation of quasimolecular orbitals (QMO) 
\cite{SinghDJ2015a,StreltsovS2015a}

\begin{figure}[b]
\includegraphics[width=0.23\textwidth]{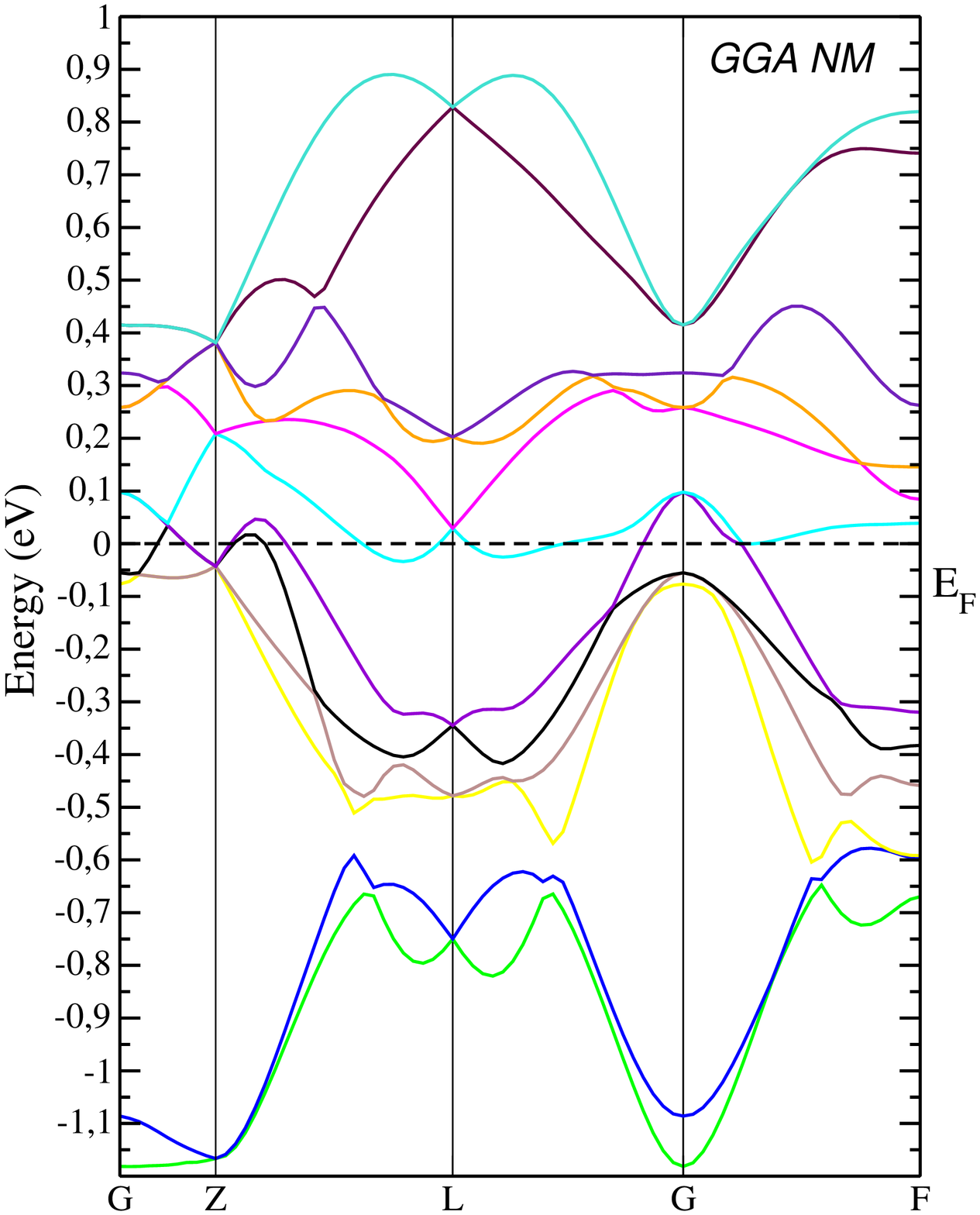}
\includegraphics[width=0.23\textwidth]{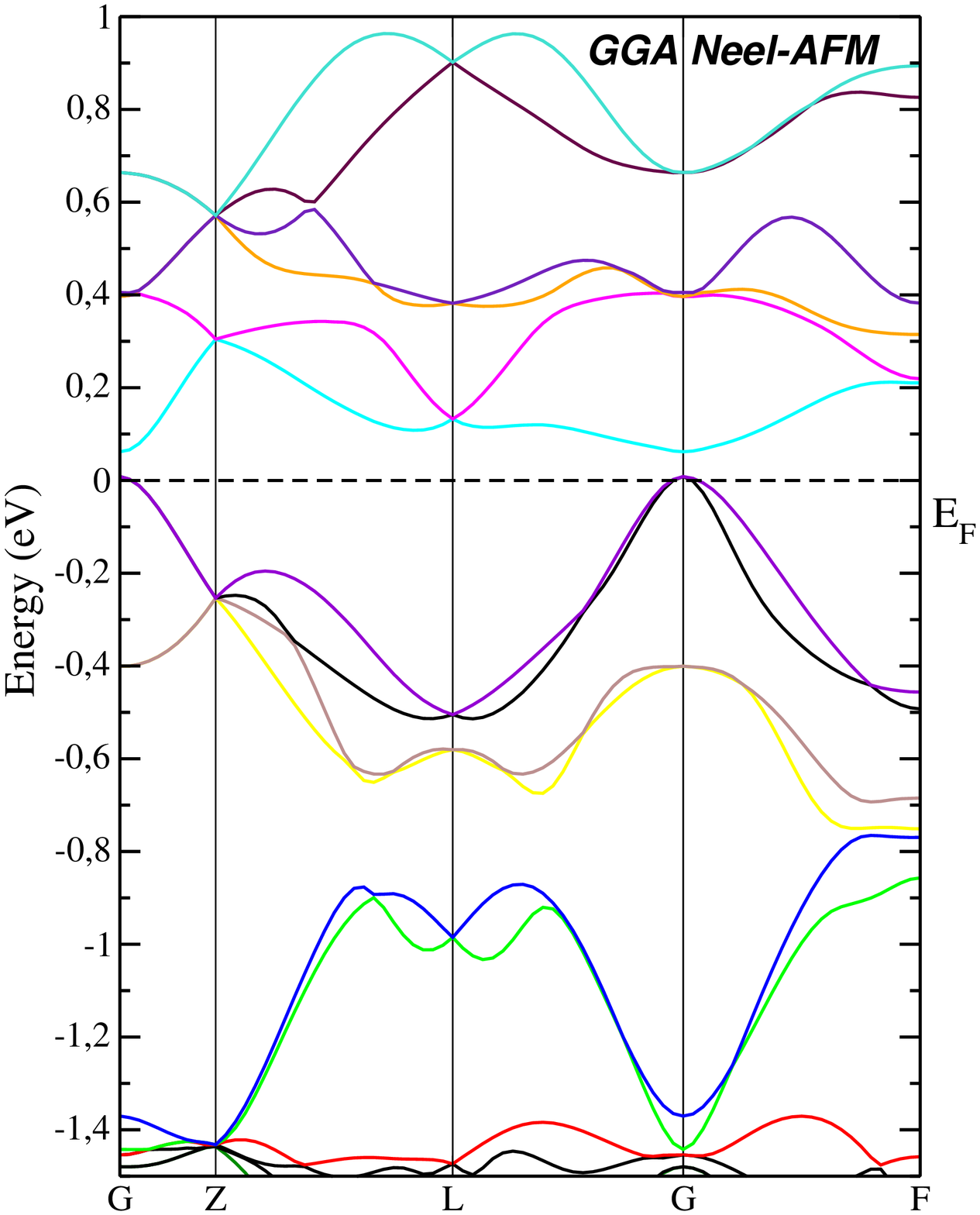}
\caption{Band structure as obtained in the GGA for nonmagnetic (left) and
N\'eel-AFM (right) configurations as calculated in Wien2k~\cite{WIEN2k}.}
\label{fig:bands} 
\end{figure}

However, there is a difference between AgRuO$_3$ and SrRu$_2$O$_6$. First of
all, in AgRuO$_3$ in the nonmagnetic state the bands derived from the $E_{2g}$
and $E_{1u}$ QMO, rendering it metallic, see left panel in Fig.\
\ref{fig:bands}, while the nonmagnetic SrRu$_2$O$_6$ is a semiconductor with
the band gap of 60 meV \cite{SinghDJ2015a}. Accounting for the spin
polarization in the (N\'eel-AFM) opens the gap of 80\,meV in AgRuO$_3$, see
right panel in Fig.\ \ref{fig:bands}. This agrees with the experimental
estimation of the activation energy of 29--188\,meV from an Arrhenius fit of
the electric resistance \cite{PrasadBE2017a}.

The overall band structures in AgRuO$_3$ and SrRu$_2$O$_6$ are similar and
again indicates the presence of QMO \cite{StreltsovS2015a,StreltsovS2018a} in
AgRuO$_3$. However, the unit cell in AgRuO$_3$ is two times larger (different
stacking along $c$) and therefore the number of the bands is twice that of
SrRu$_2$O$_6$. The lowest two bands are of $B_{1u}$ symmetry, then four
$E_{2g}$ bands and then six bands two of which are of $A_{1g}$ and four of
$E_{1u}$ symmetries.

The most important ingredient for the formation of QMO is the oxygen assisted
hopping between unlike $t_{2g}$ orbitals of nearest Ru ions, $t^{\prime}_1$
(if these Ru ions are in the $xy$ plane, then these will be $xz_1 - p_z -
yz_2$ and $yz_1 - p_z - xz_2$ hopping, where 1 and 2 are ions
indexes) \cite{MazinII2012a}. The Wannier function projection procedure shows
that $t^{\prime}_1 = 0.28$ eV. There is also direct hopping between the same
$t_{2g}$ orbitals of nearest neighbor Ru ions (for the same $xy$ plaquette
this will be a hopping between $xy$ orbitals) $t_1 = -0.27$\,eV. This is
different from SrRu$_2$O$_6$, where $t^{\prime}_1 \approx |2t_1|$ is
responsible for the formation of the QMO \cite{WangWangWang2015a}. Thus, the
QMO are weaker in AgRuO$_3$. Antiferromagnetism works against formation of QMO 
and therefore already on this stage we expect that the intraplane exchange 
interaction is stronger in AgRuO$_3$.

\begin{figure}[tbh]
\includegraphics[width=0.44\textwidth]{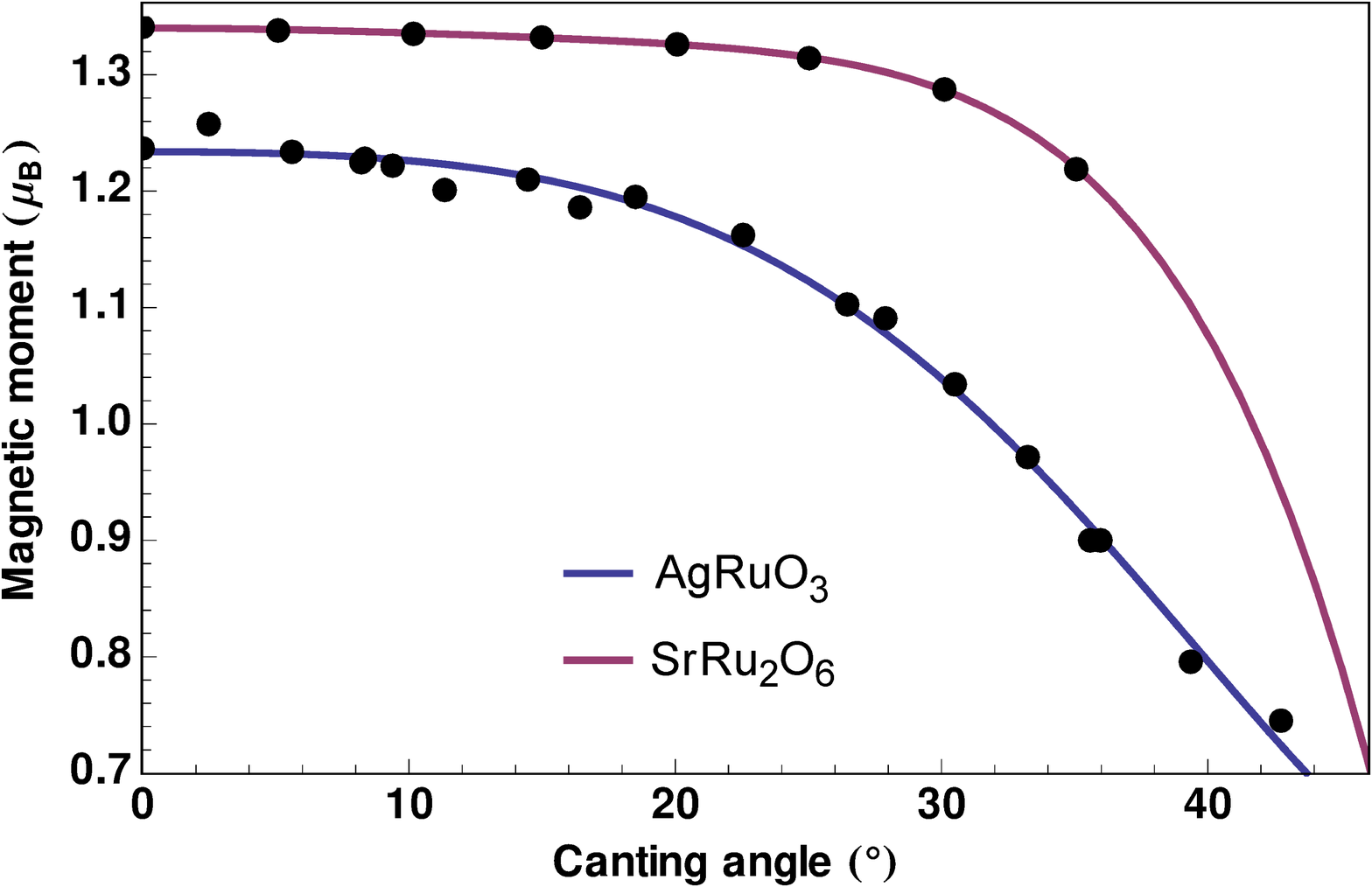}
\caption{Magnetic moment dependence on the spin canting angle ($\phi$). Spins
of two neighboring Ru ions are rotated in the plane of honeycomb lattice (so
that angle between spins is $2 \phi$). $\phi = 0$ corresponds to the N\'eel
AFM structure.}
\label{fig:ggacomp}
\end{figure}

We used the total energy GGA+SOC calculations (in the N\'eel-AFM state) to
estimate the single ion anisotropy (SIA), which turns out to be the easy axis
with $D = -12$\,K, where $D = \delta E/M_z^2$ and $M_z$ is $z$ projection of
the spin moment, $\delta E$ is the energy difference between configurations
with all spins lying in and perpendicular to the Ru-Ru plane. The
corresponding total energy dependence on the spin canting angle is shown in
Fig.\ \ref{fig:ggacomp}. We see that the SIA constant, $D$, in AgRuO$_3$
is slightly larger than in SrRu$_2$O$_6$ ($D = -9$\,K) \cite{StreltsovS2015a}.

The interlayer exchange interaction in AgRuO$_3$ is more tricky than in
SrRu$_2$O$_6$ because of different stacking. Neighboring Ru layers are shifted
with respect to each other as shown in Fig.\ \ref{fig:pnd}. As a result the
the shortest interplanar Ru-Ru bond (5.28\,{\AA}) for every Ru lead either to
the plane below or to the plane above. However, because of the N\'eel order in
the planes, the net interaction between the neighboring plane is always the
same for all planes and can be described by one net exchange constant. We have
calculated it to be $J_{\perp} = 2$\,K, which is much smaller than in
SrRu$_2$O$_6$ ($J_{\perp} = 10$\,K). The sign of the interaction is positive,
which automatically ensures the AFM stacking and the
$R\bar{3}^{\prime}c^{\prime}$ structure.

Several times smaller interplanar coupling does not affect the magnetic
properties much because it occurs logarithmically as a ration to the very
large intraplane exchange coupling. Since there is a single stable magnetic
configuration one cannot recalculate $J$ from energies of different magnetic
solutions. Therefore we used the same strategy as was previously applied to
SrRu$_2$O$_6$: we calculated in GGA+SOC the energy dependence on the spin 
canting angle and fitted the result to the functional presented in Ref.\
\cite{StreltsovS2015a}. This yields $J = 96$\,K for Hamiltonian \ref{eq:Ham} 
or 255\,K, if one uses
magnetization instead of spins as was done in \cite{StreltsovS2015a}.
Corresponding data are shown in Fig.\ \ref{fig:ggacomp}, which demonstrate
that spins are rather soft in AgRuO$_3$. If one still neglects this effect,
then the spin-wave theory can be used to estimate the N\'eel temperature
according to \cite{SchmidtB2012a,KomlevaV2020a}. With calculated intra- and
interplane exchange parameters we get $T_\mathrm{N} = 304$\,K, which is close
to experimental 342\,K. It should be mentioned that the calculated value of 
the intraplane exchange constant is also consistent with the Raman 
spectroscopy results.


\section{Conclusions}
\label{sec:conclusions}

AgRuO$_3$ is a trigonal semiconducting ruthenium oxide with Ru$^{5+}$ species
in honeycomb [Ru$_2$O$_6$] layers. It shows a N\'eel-type antiferromagnetic
order below 342(3)\,K (335--344\,K, depending on the different experimental
probes). The strongly anisotropic magnetic susceptibility (Fig.\
\ref{fig:susc}) is in agreement with the refinement of the AFM magnetic
structure from powder neutron diffraction data, as well as with first
principles calculation. The ordered Ru moments of 1.85(5)\,$\mu_\mathrm{B}$
lie (anti-)parallel to the $c$ axis (magnetic point group
$\bar{R}3^{\prime}c^{\prime}$; Fig.\ \ref{fig:pnd}), slightly larger than the
calculate moments of 1.27 $\mu_\mathrm{B}$, indicating a slightly more
correlated character of Ru $d-$orbitals. Specific heat data show a 
second-order type transition at the N\'eel temperature and no further  
transitions, in agreement with the magnetic susceptibility measurements.

The Raman and zero-field muon spin rotation spectroscopy ($\mu$SR) data indicate 
subtle changes occurring between 150\,K and 220\,K. The apparent transport gap, 
calculated as $E_\mathrm{g}(T) = T^2 d\log
\rho(T)/dT$ shows a negligible activation gap up to $\approx 40$\,K, which
than grows up to $\approx 70$\,K, reaching $\approx 25$\,meV, and then starts
to decay, reaching $\approx 0$ again at $T \sim 160$\,K. After that it grows
rapidly, reached $\approx 100$\,meV and after that remains constant up to the
N\'eel temperature. This suggests that the intrinsic semiconducting gap is
$\approx 100$\,meV, while the conductivity at low temperatures is dominated by
a small number of carriers trapped and then thermally released by some
defects.

At least one Raman mode, with the frequency $\approx 192$ cm$^{-1},$ appears
to be strongly coupled with these defects, and this coupling, as expected, is
strongly temperature dependent. While all modes, including this one, are
reproduced by the calculations, these calculations do not address the
intensity of the Raman-allowed modes; we speculate that in the ideal crystal
this mode is not visible without defect interference.

On a finer scale, one may discuss two potential energy scales. Indeed, the
changes in the Raman spectra occur at $T \approx 170$\,K. This temperature is 
roughly in the middle of the crossover region (125\,K$< T <$ 200\,K) identified 
in the ZF-$\mu$SR data, above which only a single oscillation mode is observed. 
Also, at about the same temperature the activation gap from electrical conductivity 
saturates. It is tempting to conclude that there are two energy scales, possibly 
associated with the two trapping sites, one of the order of 15--170\,K, and the 
other 200--225\,K. 

Interestingly, these defects states do not contribute to bulk properties like 
the magnetic susceptibility and specific heat. The linear term in the low-$T$ 
specific heat ($\gamma^{\prime} = 5.6(1)$ mJ mol$^{-1}$ K$^{-2}$) is sizable 
but not uncommonly large for a polycrystalline sample of an insulating oxide. 
This corroborates our picture that defects, but not changes in the basic 
electronic system of the compound, are responsible for the observed 
spectroscopic effects.


\begin{acknowledgments}
We thank S.\ Scharsach and M.\ Schmidt for the DSC measurements and 
M.\ Baenitz and C.\ Shekhar for some measurements (not shown) in the
early stage of this study. 
A.K.S.\ acknowledges the Department of Science and Technology (DST), India. 
S.P.\ acknowledges DST for an Inspire Fellowship. 
I.I.M.\ acknowledges support from the U.S.\ Department of Energy through 
grant {\#}DE-SC0021089. 
E.V.K.\ and S.V.S.\ thank the Russian Foundation for Basic Researches 
(grants 20-32-70019 and 20-32-90073) and the Russian Ministry of Science and 
High Education via program `Quantum' (grant AAAA-A18-118020190095-4) 
and contract 02.A03.21.0006. 
B.L.\ acknowledges support from the Deutsche Forschungsgemeinschaft (DFG) 
through project B06 of the SFB 1143 (ID:247310070).
This work is partially based on experiments performed at the Swiss Muon Source S$\mu$S, 
Paul Scherrer Institute, Villigen, Switzerland.
\end{acknowledgments}


%

\end{document}


\title{Supplemental Material for: \\
Magnetic and electronic ordering phenomena in the [Ru$_2$O$_6$] honeycomb lattice compound AgRuO$_3$}
\author{Walter Schnelle}
\email{walter.schnelle@cpfs.mpg.de}
\author{Beluvalli E.\ Prasad}
\altaffiliation[Now at ]{Department of Chemistry, RV Institute of Technology and Management, Bangalore, 560076, India}
\author{Claudia Felser}
\author{Martin Jansen}
\affiliation{Max Planck Institute for Chemical Physics of Solids, 01187 Dresden, Germany}
\author{Evgenia V.\ Komleva}
\author{Sergey V.\ Streltsov}
\affiliation{M.\ N.\ Miheev Institute of Metal Physics of Ural Branch of Russian Academy of Sciences, 620137 Ekaterinburg, Russia}
\affiliation{Ural Federal University, Mira St.\ 19, 620002 Ekaterinburg, Russia}
\author{Igor I.\ Mazin}
\affiliation{Department of Physics and Astronomy and Quantum Science and Engineering Center, 
George Mason University, 22030 Fairfax, Virginia, USA}
\author{Dmitry Khalyavin}
\author{Pascal Manuel}
\affiliation{ISIS Neutron and Muon Source, Rutherford Appleton Laboratory, Didcot OX11 0QX, U.K.}
\author{Sukanya Pal}
\author{D.\ V.\ S.\ Muthu}
\author{A. K.\ Sood}
\affiliation{Department of Physics, Indian Institute of Science, Bangalore 560012, India}
\author{Ekaterina S.\ Klyushina}
\author{Bella Lake}
\affiliation{Helmholtz Zentrum Berlin f\"ur Materialien und Energie, 14109 Berlin, Germany}
\affiliation{Institut f\"ur Festk\"orperphysik, Technische Universit\"at Berlin, 10623 Berlin, Germany}
\author{Jean-Christophe Orain}
\author{Hubertus Luetkens}
\affiliation{Laboratory for Muon-Spin Spectroscopy, Paul Scherrer Institute, 5232 Villigen PSI, Switzerland}
\date{\today}

\begin{abstract}
\end{abstract}

\maketitle

\setcounter{totalnumber}{4}
\setcounter{topnumber}{2}
\setcounter{bottomnumber}{2}
\setlength{\floatsep}{0ex}


\begin{table*}[p]
\caption{Experimentally observed frequencies at $T = 77$\,K and calculated 
frequencies at 0\,K along with assigned mode symmetries of the ($R\bar{3}c$) 
trigonal state of AgRuO$_3$.}
\begin{ruledtabular}
\begin{tabular}{lccc}
Mode index  & Experimentally observed (cm$^{-1}$) (77 K) & Calculated (cm$^{-1}$) (0 K) & Assigned symmetry \\
\hline
M1  &  71.2 &  70.4             & $E_{g}$          \\
M2  &  92.5 &  87.1/90.3        & $A_{1g}$/$E_{g}$ \\
M3  & 175.9 & 179.0             & $A_{1g}$         \\
M4  & 193.0 & 198.1             & $E_{g}$          \\
M5  & 200.0 & 202.2             & $E_{g}$          \\
M6  & 311.4 & 303.7/309.1       & $E_{g}$          \\
M7  & 332.5 & 344.0             & $A_{1g}$         \\
M8  & 480.0 & 478.6             & $E_{g}$          \\
M9  & 495.0 & 484.1/486.2/499.7 & $E_{g}$          \\
M10 & 526.0 & 527.9             & $E_{g}$          \\
M11 & 552.3 & 557.3             & $A_{1g}$         \\
M12 & 595.8 & 579.5/580.1       & $A_{1}$/$A_{2u}$ \\
\end{tabular}
\end{ruledtabular}
\label{tab:raman}
\end{table*}

\begin{figure}[p]
\includegraphics[width=0.48\textwidth]{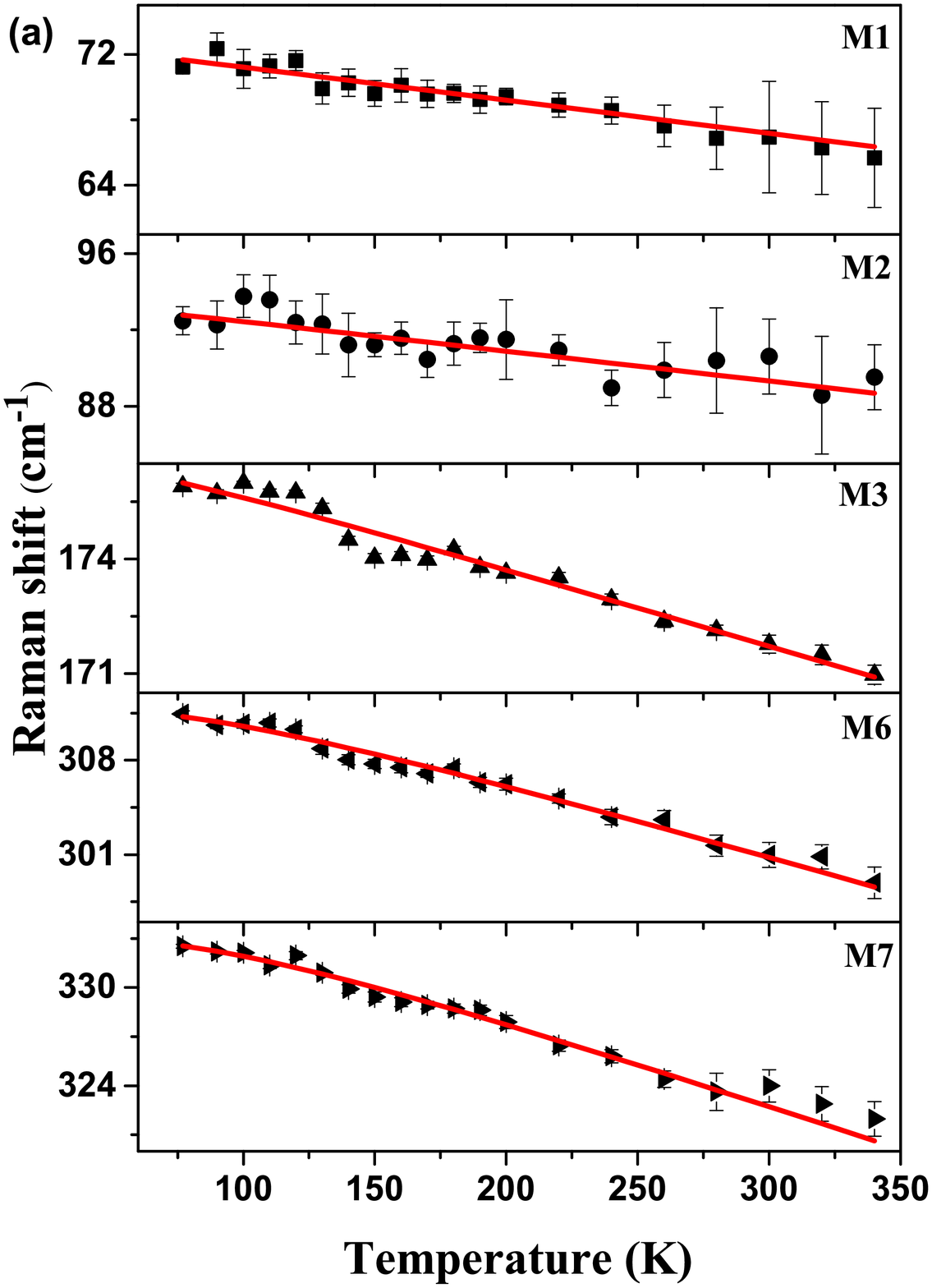}
\includegraphics[width=0.48\textwidth]{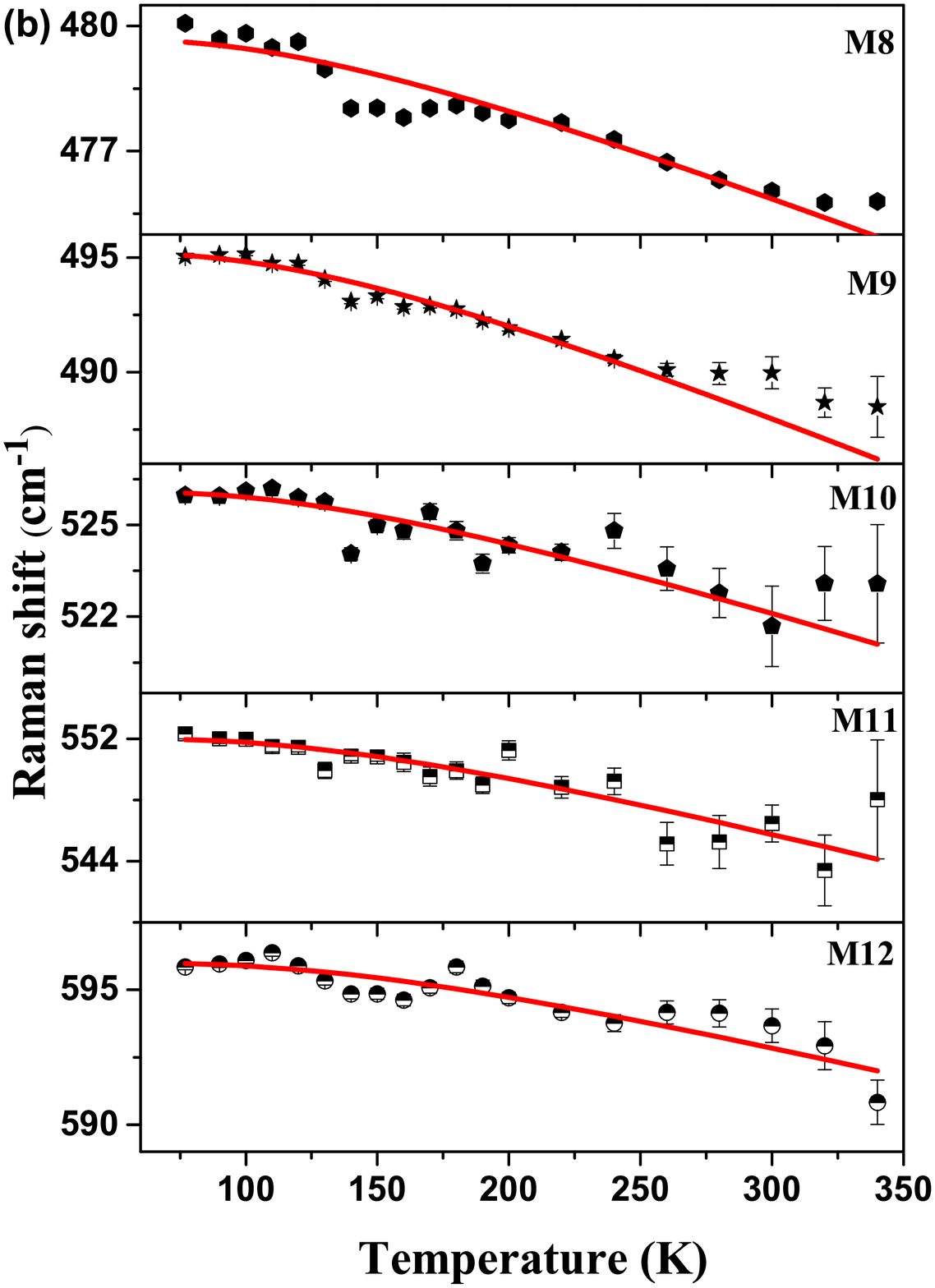}
\caption{Temperature dependence of phonon frequencies of the Raman modes of
AgRuO$_3$. The error bars are also displayed and are less than the size of the
symbol when not shown. The red solid lines are fit to a simple cubic
anharmonic model to the experimental data.}
\label{fig:raman3}
\end{figure}

\begin{figure}[p]
\includegraphics[width=0.48\textwidth]{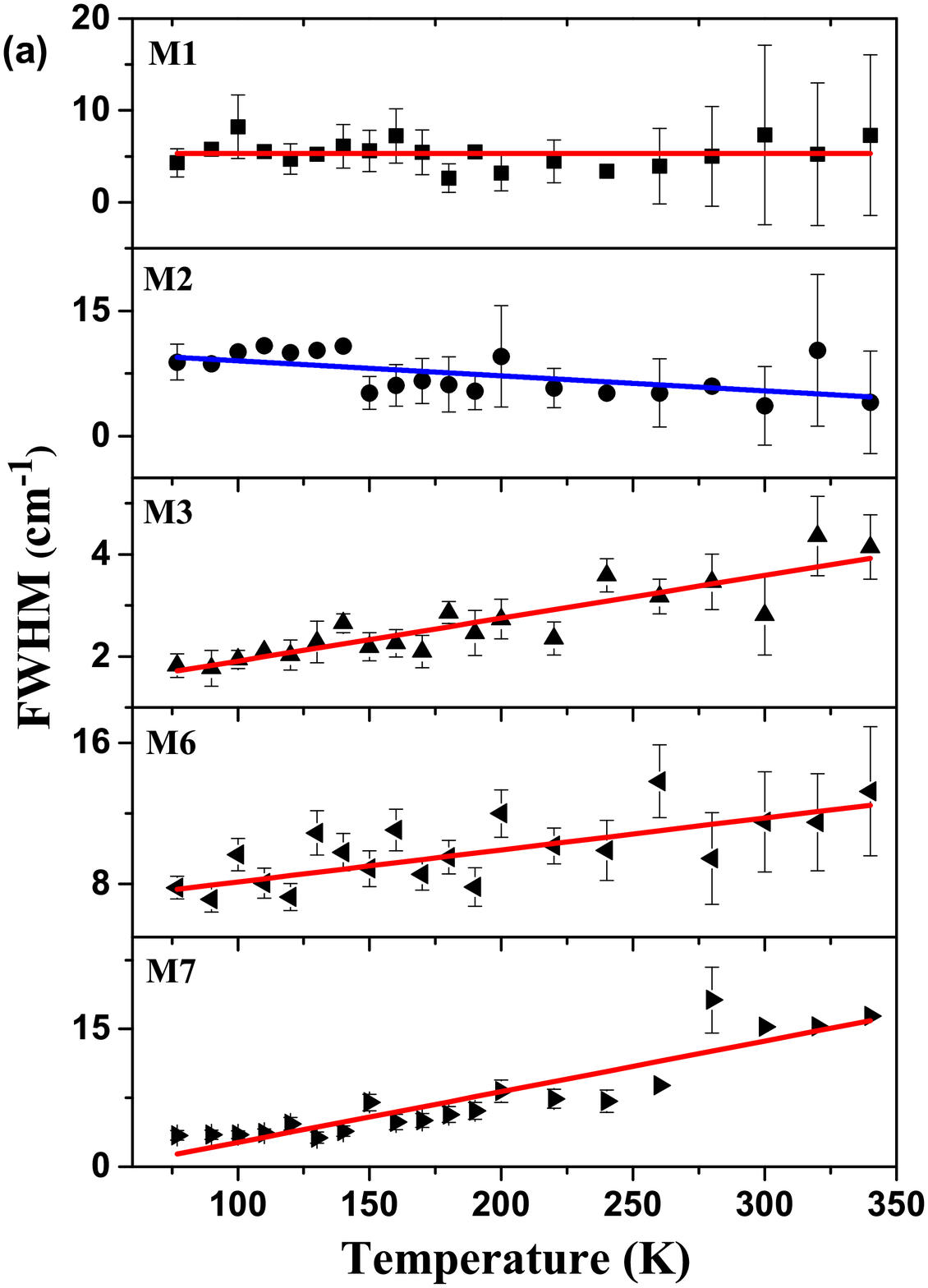}
\includegraphics[width=0.48\textwidth]{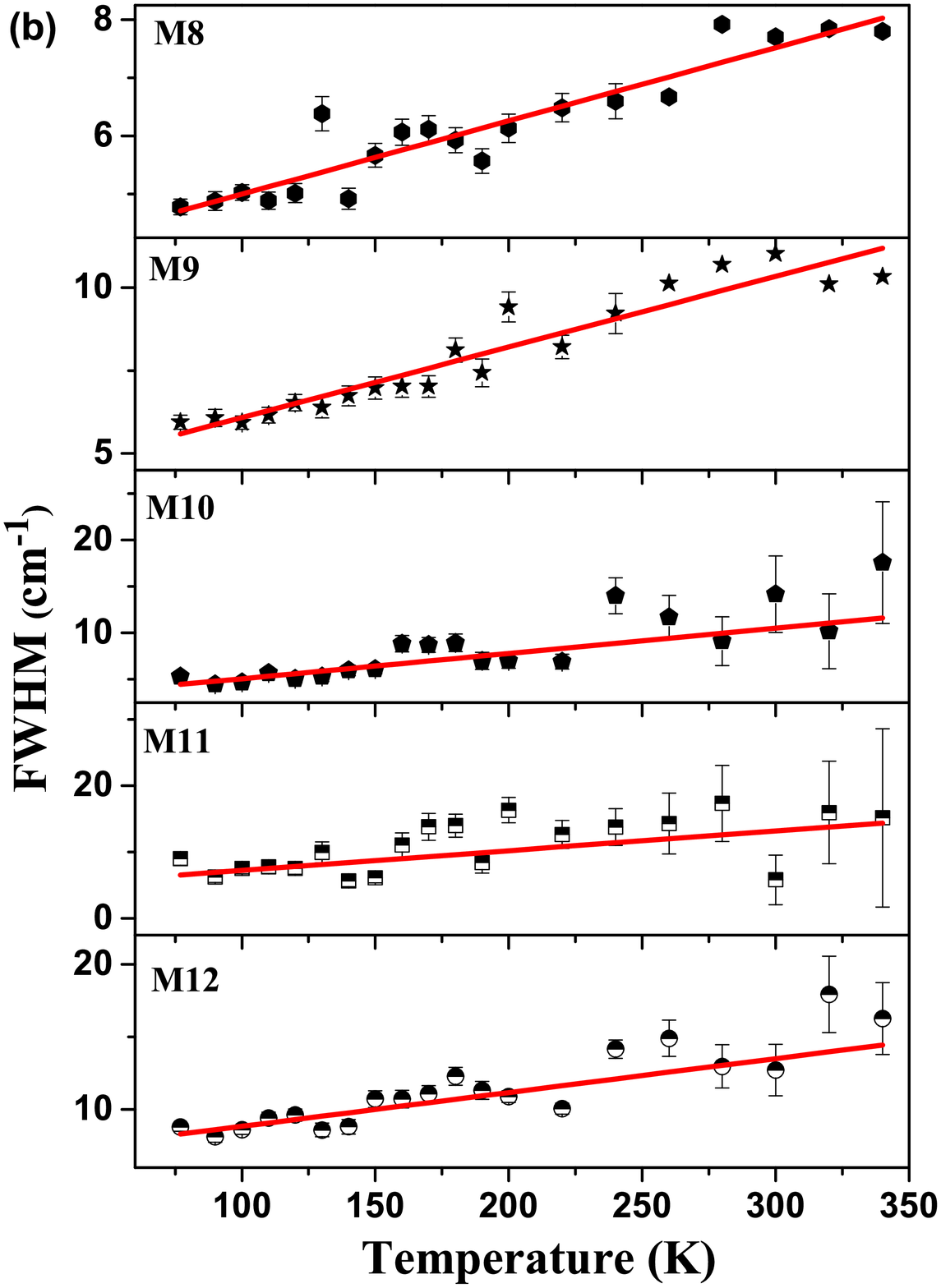}
\caption{Temperature dependence of phonon linewidths of the Raman modes of
AgRuO$_3$. The red solid lines are fit to a simple cubic anharmonic model to
the experimental data. FWHM of the mode M2 remains almost
constant and the solid blue line linear fit to the data.}
\label{fig:raman4}
\end{figure}

\begin{figure}[p]
\includegraphics[width=0.53\textwidth]{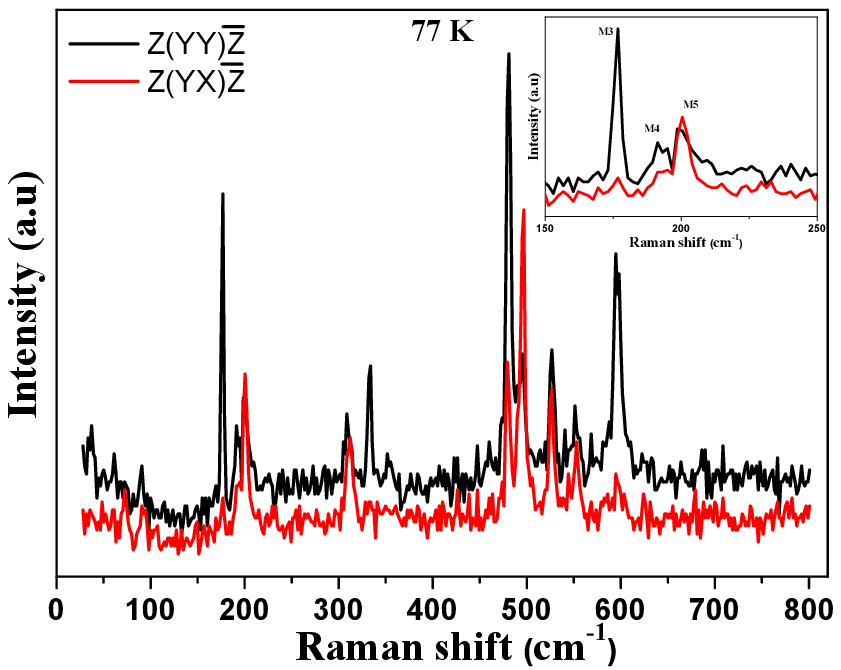}
\caption{Raman spectra of AgRuO$_3$ measured at $T = 77$\,K in different polarization 
geometries. The inset shows zoomed-in spectra in the range 150--200\,$cm^{-1}$ 
confirming the presence of modes M4 and M5 in both parallel and crossed polarization 
configurations.}
\label{fig:ramanpol}
\end{figure}